\documentstyle[12pt,a4wide,epsf,epsfig]{article}

\begin{document} 

\begin{titlepage} 

\begin{flushright}
FTUV-03-0702 \\
\end{flushright}

\vspace*{2cm} 

\begin{centering}

{\Large {\bf Super Heavy Dark Matter Anisotropies \\
from D-particles in the Early Universe}}

\vspace*{1cm} 

{\bf N. E. Mavromatos}$^{a,b}$ and {\bf J. Papavassiliou}$^{b}$

\vspace*{0.2cm} 

$^{a}$ King's College London, Dept.of Physics (Theoretical Physics), 
Strand WC2R 2LS, U.K.\\

$^{b}$ Departamento de F\'\i sica Te\'orica and IFIC, 
Universidad de Valencia-CSIC, E-46100, Burjassot, Valencia, Spain.

\vspace*{1.5cm} 

{\bf Abstract} 

\end{centering} 

\vspace*{0.2cm} 

{\small We discuss a way  of producing anisotropies in the spectrum of
superheavy  Dark  matter, which  are  due  to  the distortion  of  the
inflationary  space time  induced by  the recoil  of  D-particles upon
their scattering  with ordinary string  matter in the  Early Universe.
We calculate  such distortions by world-sheet  Liouville string theory
(perturbative) methods.   The resulting  anisotropies are found  to be
proportional  to  the  average  recoil  velocity and  density  of  the
D-particles.   In our  analysis we  employ a  regulated version  of de
Sitter  space,  allowing  for   graceful  exit  from  inflation.  This
guarantees the asymptotic flatness of  the space time, as required for
a consistent interpretation, within an effective field theory context,
of   the  associated   Bogolubov  coefficients   as   particle  number
densities. The latter are computed by standard WKB methods.}

\vspace{4cm}

\begin{centering}

{\it Dedicated to the memory of Ian I. Kogan, collaborator and friend}

\end{centering}

\end{titlepage} 

\newpage

\section{Introduction and Summary}

A plethora of recent  astrophysical data, ranging from direct evidence
on   the   acceleration   of   the  Universe   using   supernovae   Ia
data~\cite{supernovae}  to   measurements  of  the   cosmic  microwave
background  anisotropies  (CMB)   to  an  unprecedented  precision  by
WMAP~\cite{wmap}, support  strongly two important  characteristics for
our       observable      Universe~\cite{bellido}:       (i)      {\it
inflation}~\cite{inflation}, i.e.  a  phase with an exponentially expanding
scale  factor  in the  Robertson-Walker  space-time  seems  to be  an
integral  component of  the (early)  evolution of  our  Universe, (ii)
$70\%$  of the  Universe  energy  content consists  of  a yet  unknown
substance, termed {\it dark energy}.

Inflationary  dynamics is  supported by  the spatial  flatness  of the
Universe, corroborated by the  CMB  data. In  the standard  field
theoretic  implementation such  a phase  requires the  presence  of a
scalar field  mode, the  {\it inflaton} field,  whose nature  is still
unknown.  The WMAP  data  are still  too crude~\cite{wmapinflaton}  to
determine the precise shape of its potential.  The dark energy, on the
other hand,  may be  either a cosmological  {\it constant}, or  a {\it
relaxing  to  zero}  component  of  the  `vacuum'  energy,  due  to  a
non-equilibrium     situation,     e.g.     a    {\it     quintessence
field}~\cite{carroll}, or  in general some excitation  of our Universe
due to  an initial cosmically  catastrophic event.  WMAP  has measured
the equation of  state $p = w\rho$, $p$ being  the pressure and $\rho$
the energy  density, of  such a  quintessence field, and  found $  w <
-0.78$ (for comparison the cosmological constant model has $w = -1$).

In the modern context of string theory~\cite{strings}, the presence of
a cosmological constant (de  Sitter Universe) is an unwelcome feature,
due to the complications produced by the existence of an event horizon
when  attempting  to define  proper  asymptotic  states,  and thus  an
S-matrix~\cite{smatrix}.   Indeed, a  non-trivial but  constant vacuum
energy  density in  a  Friedman-Robertson-Walker (standard)  cosmology
will  eventually dominate the  evolution of  the Universe,  which will
re-enter  into an  accelerating  inflationary phase,  thus failing  to
reach  an asymptotically  flat domain.  Given that  string  theory, at
least as we  understand it today, is based  on S-matrix elements, such
backgrounds appear problematic from this point of view.  On the other
hand, relaxation  or quintessence-like scenarios,  despite fine-tuning
drawbacks related to  the shape of the scalar-mode potentials, allow 
the possibility of an asymptotic S-matrix, and hence they 
may be solutions of some versions of string theory.

Such a situation has been discussed in the context of string theory in
ref.~\cite{papant}, and in the modern context of brane
cosmology~\cite{branecosm} in ref.~\cite{grav2},  in a model involving
colliding brane  worlds. In addition, colliding-world  scenaria of the
``ekpyrotic''  type  have  been  discussed in  the  recent  literature
~\cite{ekpyrotic}, where  it was suggested that  an inflationary phase
were  absent and unnecessary.   This point  of view  appears to  be in
direct   conflict  with  the   above-mentioned  recent   evidence  for
inflation.  Moreover,  this approach has been criticized  in a stringy
context~\cite{linde},  by arguing that  classical string  equations of
motion  (conformal invariance  conditions)  do not  lead to  expanding
Universes but rather to contracting ones~\footnote{This last point has
been refuted, however,  in \cite{seiberg} based on the  existence of a
hypothetical (non-perturbative) stringy phase-transition.}.
 
In  this respect,  a different  point of  view has  been  advocated in
\cite{grav2}, where the collision of  the brane worlds has been viewed
as a  {\it non-equilibrium stringy process}, quantified  within a {\it
non-critical}   (Liouville)  string  theory~\cite{ddk,emn}   upon  the
identification     of    target     time     with    the     Liouville
mode~\cite{kogan1,emn}.   The   `catastrophic  cosmic  event'   of  the
collision of the brane worlds leads to a central charge deficit in the
pertinent $\sigma$-model,  which describes  in a perturbative  way the
stringy excitations of our (brane) Universe after the collision.

An  important  consequence  of  this departure  from  critical  string
theory,  and thus  from the  standard conformal  invariance conditions
used in  \cite{linde}, is the  presence of an  exponentially expanding
{\it    inflationary}   phase    for   the    four-dimensional   scale
factor. Moreover,  such models, asymptotically  lead to a  current era
with a relaxing to zero quintessence-like dark energy component of the
Universe, scaling with the cosmic time as $1/t^2$~\footnote{We note in
passing that  in the model of \cite{grav2}  such relaxing-to-zero dark
energy  component at  the  current  era can  be  made compatible  with
standard  supersymmetry breaking  models, with  the  symmetry breaking
scale within the  range of a few TeV.}.  It is  important to notice that
such a  scaling is computed  using (logarithmic~\cite{lcft}) conformal
field theory  methods.  The inflationary  as well as  the accelerating
(late)  phases  of the  Universe  in  such  models occur  dynamically,
without  the   introduction  of  extra  scalar   fields  (inflaton  or
quintessence).  In non-critical string theory such inflationary phases
are obtained~\cite{emninfl} upon the identification of the target time
with  the Liouville  mode~\cite{kogan1,emn}; the  consistency  of this
procedure  has  been checked  in  several  models  so far.  Here  this
approach is reviewed in the Appendix, for the colliding world scenario
of \cite{grav2}.

In  this work, we  adopt the  above scenario,  and proceed  further to
discuss  particle  production  during  the  inflationary  era,  in  an
extended  brane model,  where the  three-dimensional brane  worlds are
punctured by  D-particles~\cite{strings}. This situation  is viewed as
the simplest  case of intersecting  branes; it may occur  naturally in
M-theory scenaria,  where the branes are  viewed as defects  in a bulk
space-time.   It is  the purpose  of this  article to  argue  that the
presence of D-particles on  the brane worlds~\footnote{We note that in
general  the density of  $D$ particles  in our  brane Universe  may be
bounded from above  by the requirement that their  presence should not
affect  the current  upper limits  on  the value  of the  cosmological
constant, which stem from  the requirement of the large-scale validity
of Newtonian  mechanics.  However,  supersymmetry in target  space may
play a subtle r\^ole, resulting  in zero ground state energies, should
one view the punctured (by  D-particles) three-brane as a ground state
of  some  (initially)  supersymmetric  M-theory  configuration.   Such
issues, however, fall far beyond  the scope of the present article and
will not  be discussed  further here.}  leads  to anisotropies  in the
spectrum  of the heavy  particles emitted.   Such anisotropies  can be
cosmologically relevant if the masses  of the emitted particles are of
order $10^{13 \pm 4}$ GeV~\cite{chung}.

In view of scenaria~\cite{kryptons} in which such heavy particles 
may be the sources of ultra-high-energy cosmic rays, the present model 
provides a way of obtaining anisotropies in their spectra. 
Although at present there is no experimental evidence for 
such anisotropies~\cite{uhecranis}, the situation 
is far from being conclusive, due to the scarcity  
of the available ultra-high-energy cosmic ray events. In fact, it is expected
that in the near future the experimental precision will improve 
significantly. 
Should anisotropies be eventually detected, 
the present model 
could provide a possible explanation for their origin.
In the contrary situation, 
since such anisotropies constitute
a concrete prediction of the current model,
one will be able of placing strict bounds on the density  
of (primordial) D-particles on our brane Universe, 
a parameter of central importance in this specific  model.  
Moreover, this model seems realistic from the M-theory point of view,
where intersecting brane situations might be a natural 
possibility, with interesting cosmological implications~\cite{intersect},
and hence we consider it as worthy of being explored further.

The structure  of the article is  as follows: in section  2 we present
the non-critical string formalism describing the recoil of D-particles
in a brane Universe, due to scattering of string matter propagating on
the brane.   In sections  3 and 4  we use  world-sheet renormalization
group methods to compute the back-reaction effects on the inflationary
space time  Geometry from a  recoiling D-particle.  We  pay particular
attention  to discussing  the  range of  parameters  that guarantee  a
perturbative  $\sigma$-model treatment.  Specifically, we  can compute
such effects via conformal field theory  methods only for times $0 < t
<< 1/H$, where  $H$ is the Hubble parameter  (assumed constant) during
inflation. As  a consistency check of  the approach, we  find that the
back-reaction  distortion  indeed  attenuates  exponentially  as  time
progresses.   In  section 5  we  discuss  particle  production in  the
inflationary    universe,    due   to    the    presence   of    these
D-particle-recoil-induced   distortions.    Using  approximate   (WKB)
methods~\cite{chung}, standard  in such  kind of problems,  we compute
the  associated  massive  particle  number  density by  means  of  the
relevant Bogolubov coefficient.  In  section 6 we discuss the particle
spectrum  so obtained.   We find  that  the effect  of the  space-time
distortion induced by the D-particle  is to make the momentum spectrum
for  massive  particles non-Gaussian,  and  non-thermal, with  angular
dependence leading to anisotropies.   The estimates of \cite{chung} on
the  mass of  such particles  $M >  10^{13}$ GeV,  in order  for their
density to be  cosmologically relevant, and thus to  act as superheavy
dark matter candidates, are still valid in our case.  The anisotropies
we  find   in  the   production  spectrum  will   imply  corresponding
anisotropies in the ultra-high-energy  cosmic ray spectrum, should the
latter  be  produced  by  such  heavy  particles,  according  to  some
scenaria~\cite{kryptons}.   Conclusions  and   outlook  are  given  in
section  7. Finally,  in an  Appendix we  review briefly  how  one can
obtain   inflationary   space-times   as  consistent   $\sigma$-model
backgrounds of a non-critical string model.

\section{Geometry fluctuations due to D-brane recoil: Non-critical String
Formalism}

Consider the single scattering event of a closed string with a D-particle
defect embedded in a $D$-dimensional space time of metric $G_{\mu\nu}$.
Long after the scattering the induced disturbance of the neighboring space time
(around the recoiling defect) is determined by world-sheet methods~\cite{kogan,grav}.
Specifically, the recoil of a D-particle, under its scattering off 
a (closed)
string state in a metric background $G_{\mu\nu}$ 
is determined, at a $\sigma$-model level, by the 
deformation vertex operator~\cite{grav}:
\begin{equation} \label{path}
V = \int _{\partial \Sigma} G_{ij}y^j(t)\Theta_\epsilon (t) 
\partial_n X^i 
\end{equation}
with $\Theta_\epsilon (t)$ 
the regularized Heaviside `impulse' operator:
\begin{equation}
\Theta_\epsilon (t) \sim 
\frac{1}{i}\int_{-\infty}^{+\infty}\frac{d\omega}{\omega} e^{i\omega t}
\label{regulator}
\end{equation}
where $\epsilon \to 0^+$, $G_{ij}$ 
denotes the spatial components of the metric, 
$\partial \Sigma$ denotes the world-sheet boundary, 
$\partial _n$ is a world-sheet derivative 
normal to the boundary,
$X^i$ are $\sigma$-model fields obeying Dirichlet boundary conditions 
on the world sheet, 
and $t$ is a $\sigma$-model field obeying Neumann boundary conditions
on the world sheet, whose zero mode is the target time. 
The quantity $y^i(t)$, $i$ being a spatial index,   
denotes the trajectory of the $D$-particle. 

This is the basic vertex deformation, which we assume 
to describe the motion of a $D$-particle in a curved geometry at least   
to leading order, when space-time back reaction 
and curvature effects are assumed weak~\footnote{Perhaps a formally more desirable approach 
towards the construction of the complete vertex operator 
would be to start from 
a T-dual (Neumann) picture, where the deformation (\ref{path}) 
should correspond to a proper Wilson loop operator of an
appropriate  
gauge vector field. Such loop operators are by construction independent
of the background geometry. One can 
then pass onto the Dirichlet picture by a T-duality transformation
viewed as a canonical transformation from a $\sigma$-model 
viewpoint~\cite{otto}. In principle, such a procedure  
would yield a complete form of the vertex operator in the Dirichlet 
picture, 
describing the path of a $D$-particle in a curved geometry.
Unfortunately, such a procedure is not free from ambiguities
at a quantum level~\cite{otto}, 
which are still unresolved for general curved backgrounds.}.
Such vertex deformations may be viewed as a 
generalization of the flat-target-space case~\cite{kogan}.

For times relatively long after the event, 
where the use of $\sigma$-model perturbation theory 
is a valid approximation,   
the trajectory $y^i(t)$ will be that of free motion
in the 
curved space-time under consideration. In the flat space-time 
case, this trajectory was a straight line~\cite{kogan}, 
and in the more general 
case here it will be simply the associated {\it geodesic}. 
Let us now determine its form, which will be essential in what follows. 

For our purposes we shall consider {\it Minkowski signature} 
space-time backgrounds
of Robertson-Walker (RW) form: 
\begin{equation}\label{rwmetric}
ds^2 = -dt^2 + a(t)^2 (dX^i)^2  
\end{equation}
where $a(t)$ is the RW scale factor. 

An important discussion is in due order at this point. In order for 
the basic conformal field theory formalism to be applicable
in curved target space-times, which has been used in \cite{grav}, it is 
{\it imperative} that these  
space-times be {\it at least an approximate}
solution to $\sigma$-model conformal invariant conditions (i.e. a vacuum solution to Einstein's equations to leading order in $\alpha '$ expansion,
where $\alpha ' = 1/M_s^2$ is the Regge slope, and $M_s$ is the string scale). 
This is the case for RW backgrounds at large times $ t \to \infty$,  
presented in \cite{grav}, where target space
curvature effects have been ignored. 
In the present paper we are interested in inflationary metric backgrounds.
Unfortunately, such backgrounds are not vacuum solutions to any string theory,
at least in an `unregulated' form, because in this case they are hampered
by event cosmological horizons, and hence are not compatible 
with a consistent $S$-matrix. The latter is a cornerstone of 
the first-quantized formulation of string theory, and hence consistent 
string theory backgrounds should allow for a graceful exit 
from an inflationary phase, and approach asymptotically a solution of the 
conformal invariance $\sigma$-model conditions. Such situations have 
been argued to characterize realistic non-critical 
string cosmologies under the identification of time with the 
Liouville mode~\cite{papant}. As mentioned previously, for our purposes 
in this work we discuss briefly 
such a case in the Appendix, specifically the colliding brane
model of \cite{grav2}.

Once we assume such regulated versions of de Sitter space-time, 
we may proceed by adopting the same spirit as in \cite{grav}, 
i.e. concentrate in regions of time for which one may safely 
neglect space-time curvature effects and thus apply 
standard conformal field theory world-sheet methods.
Contrary 
to the case studied in \cite{grav2}, 
in the inflationary situation, studied here, 
curvature effects are weak compared to $G_{\mu\nu}$
in the time regime $0 \ll t H \ll 1$, where the 
event of the collision of the string with the D-particle 
is placed at $t=0$.
We stress once more  
that the left inequality is necessary for the validity of the boundary 
world-sheet $\sigma$-model perturbation theory~\cite{kogan}.

The geodesics of a massive point particle in de Sitter space
($a(t)=a_0e^{Ht}$, $a_0$ a constant whose size will be fixed later on) 
are given by: 
\begin{equation}
y^i (t) = c_3^i -\frac{c_1^i}{|c_1|H}
\left( a_0^{-2}e^{-2Ht} + \frac{2c_2}{|c_1|^2}\right)^{1/2}
\label{geodesic}
\end{equation} 
where $c_\alpha^i$, $\alpha = 1,2,3$ are constants determined by the 
boundary conditions. 

We shall be interested in values of $H \sim 10^{-5}M_P$, where 
$M_P \sim 10^{19}$ GeV, is the four dimensional 
Planck scale, not to be confused with the 
string scale $M_s $ which may be different.
This is the value of the Hubble parameter 
during the inflation era. For the validity of our perturbative 
analysis we shall also be interested 
in times far from 
the end of inflation $t \ll t_e \sim 10^{9}t_P$ (where $t_P\sim 10^{-43} 
{\rm sec}$ 
is the four-dimensional Planck time),
such that  $Ht \ll 1$, 
where space-time curvature effects of order $H^{2} \sim 10^{-10}M_P^2$ 
can be safely neglected.
However, as we shall discuss in the next two sections, we shall be 
interested in ranges of the parameters of our cosmological model 
such that 
\begin{equation} 
1 \gg \frac{H^2}{M_s^2} \gg Ht > 0~.
\label{relconstr}
\end{equation}  
It is terms of order $H^2/M_s^2$ that will determine the distortion 
of the inflationary space time due to the recoil of the D-particles. 
We easily see that a natural set of parameters satisfying (\ref{relconstr}) 
is $M_s = 10^{-4} M_P \sim 10^{15}$ GeV (i.e. intermediate string scale),
which implies that our perturbative treatment in this article is 
rigorously valid for times $t < 10^{3}t_P$ (beginning of inflationary 
period).  

This will be a sufficient,
and mathematically self-consistent, procedure for 
computing the global effects 
of the recoil-induced distortion of space-time,
which, as we shall discuss in the next section, depend on the world-sheet 
violations of conformal invariance, quantified by 
the 
anomalous dimension of some operators of order $H^2$ (in string $M_s=1$ units) 
which therefore should be kept in our analysis. 
As we shall show, Such space-time distortion effects 
diminish with time as $e^{-4Ht}$,
and hence become even more suppressed toward the end of 
inflation (where $Ht_e \sim 10^{4}$).

Working in the above regime of 
parameters, therefore, will guarantee : (i) the validity of the $\sigma$-model 
perturbation theory; (ii) the presence   
of a {\it smooth extension} of de Sitter
space, such that asymptotically one can define flat space-time regions
and a proper $S$-matrix. The latter requirement 
is necessary for a proper definition of
particle production number via the Bogolubov 
coefficients~\cite{birrell,chung,biswas,schwarz}; it may be satisfied 
{\it provided } that the constant $a_0 \ll 1$, for instance 
of order $1/Ht_e$ in Planck units. Physically, this last
condition on  $a_0$ implies that the inflationary era started at $t=0$ 
(when the recoil occurred), 
at which moment the spatial size of the  universe $a_0$ was 
considerably smaller than the Planck length. This situation corresponds 
physically to a smoothened version of 
the `big bang singularity', which we adopt for our purposes here.
It is worth stressing that in this way one may apply world-sheet 
perturbation theory at the early stages of inflation, viewing the 
latter~\cite{emninfl} as
an example of  non-critical (Liouville) $\sigma$-model~\cite{ddk},
with the identification of time with (an appropriate function) of the 
Liouville mode~\cite{kogan1,emn}.

We next remark that
the non-relativistic nature of the D-particles, appropriate 
for the validity of our $\sigma$-model formalism~\cite{kogan,grav}
imply small initial ($t=0$) recoil velocities, 
defined by: $u^i = \frac{dy^i}{dt}|_{t=0}$, which 
yields (from eq. (\ref{geodesic})):  
\begin{equation} 
u^i = \frac{c_1^i}{|c_1|a_0^2}\left(a_0^{-2} + 2\frac{c_2}{|c_1|^2}\right)^{-1/2} 
\ll 1
\end{equation} 
As mentioned earlier, 
in the regime of parameters we are 
working, one must have $a_0 \ll 1$, in which case non relativistic velocities
$u^i \ll 1$  
are guaranteed by the condition 
\begin{equation}
2a_0^{4}c_2 \gg |c_1|^2,
\label{velocities}
\end{equation}
which, in turn, implies 
\begin{equation}
u^i \simeq \frac{c_1^i}{a_0^2\sqrt{2c_2}}
\label{nonrelvel}
\end{equation}

An additional requirement, not usually imposed 
in standard inflationary scenaria of cosmology~\cite{birrell}, 
is that of  the {\it continuity} of the various results as $H \to 0$.
As will be explained in the next section, this last requirement 
is motivated by the fact that,  
during  the exponential expansion,   
the Hubble parameter $H$
acts as an anomalous dimension of world-sheet operators.
To be specific, in the Liouville treatment
we shall impose {\it finiteness} of results as $H \to 0^{+}$, in order to 
guarantee a connection with the (world-sheet RG fixed point of) flat space-time
with metric $ds_*^2 = -(d\varphi)^2 -(dt)^2 + a_0^2 (dX^i)^2$ 
(with $\varphi =-t$)~\footnote{Notice that the Liouville
mode always exist in a $\sigma$-model path integral, but decouples 
from the background fields at the fixed points. However, as a target-space 
coordinate is 
always there, and thus contributes 
an overall normalization factor of 2 to the time component of the 
flat fixed point metric $ds_*^2$; this guarantees the smoothness of the 
limit $H \to 0^+$.}. 
In fact, in this approach we shall regard the inflation as a smooth 
interpolating RG trajectory between two asymptotically flat regimes
of space-time, which are fixed points of the world-sheet of the 
string. What differentiate these two regimes is the size of $a_0$.
We shall return to this point later, when
we discuss the interpolating metric between such fixed points,
including an inflationary era, followed by a ``graceful exit'' from it,
so that an S-matrix can be properly defined, as expected in a 
string theory reaching its critical status asymptotically in time.
 
Requiring finiteness of $y^i(t)$ in the limit $H \to 0^+$ 
is equivalent to setting 
\begin{equation} 
c_3^i = \frac{c_1^i}{|c_1|H}\left( a_0^{-2} + \frac{2c_2}{|c_1|^2}\right)
+ H-{\rm independent~parts}~.
\label{hindependence}
\end{equation} 
For simplicity we assume the $H$-independent parts to be zero, 
which essentially amounts to choosing the initial position of the 
D-particle to be the origin of the coordinates and of our (inflationary) 
Universe. These considerations imply 
the following approximation for the geodesic (for the time $t$ 
regime $Ht \ll 1$ we are interested in):
\begin{equation}
y^i (t) \simeq \frac{c_1^i}{2a_0^{2}H\sqrt{2c_2}}
-\frac{c_1^i}{2a_0^{2}H\sqrt{2c_2}}e^{-2Ht} 
\label{approxgeo}
\end{equation}
Thus the recoil operators for the D-particle (\ref{path}) 
can be split in two types of $\sigma$-model deformations,
assuming that the moment of impact of the closed string state with the 
D-particle defect takes plays at $t=0$:
\begin{eqnarray}
&& {\cal C}={\tilde c}_i \oint_{\partial \Sigma} \Theta_\epsilon (t) 
\partial_n X^i \nonumber \\
&& {\cal D}={c}_{0,i} \oint_{\partial \Sigma} \Theta_\epsilon (t) e^{2Ht}
\partial_n X^i 
\label{recoilops}
\end{eqnarray}
where ${\tilde c}^i = - c_{0,i} 
\equiv -\frac{c_1^i}{2\sqrt{2c_2} H} = -a_0^2\frac{u^i}{2H}$.
Notice that, 
in general, the validity of our weak coupling $\sigma$-model formalism
requires ${\tilde c}^i, c_0^i < 1$, 
which in view of (\ref{velocities}) is satisfied.

\section{Renormalization Group Relevance of the recoil 
Deformations: Operator Product Expansion Analysis}

The Liouville approach to recoil-induced space-time distortions~\cite{kogan,emn,grav}
require the relevance of the recoil operators (\ref{recoilops}) 
under world-sheet renormalization group (RG) transformations; in turn, this  
would imply the need for ``Liouville dressing''.
The fact that the recoil operators are indeed RG-relevant 
can be confirmed by considering their 
Operator Product Expansion (O.P.E.) 
with the stress tensor of the $\sigma$-model, as well as among themselves. 

Working with time regimes $0 \ll Ht \ll 1$, and also with parameters 
satisfying (\ref{relconstr}), 
which, 
as discussed above, guarantee the validity of our perturbative $\sigma$-model method, 
justifies the use of 
the following formula for the world-sheet propagator of the 
$X^\mu = (t, X^i)$, $i=1, \dots D-1$ (spatial index) coordinate fields
(from now on we shall be working in string units for which $M_s = 1$):
\begin{equation}
<X^\mu (z) X^\nu (z') > \simeq  G^{\mu\nu}{\rm ln}|z - z'|  \;
\label{propagator}
\end{equation}
$G^{\mu\nu}$ is the target-space metric in the $\sigma$-model frame,
which is assumed of the form (\ref{rwmetric}) with a de-Sitter type 
exponentially expanding scale factor $a(t) = a_0e^{Ht}$. 
The (approximate) 
formula (\ref{propagator}) will be subsequently used when computing the various 
OPE of the $\sigma$-model operators. 

To begin with, the stress tensor assumes the form:
\begin{equation}
2T = -(\partial_z t(z))^2 + a^2(t) (\partial_z X^i)^2~; \quad 
a(t) = a_0e^{Ht}
\label{stress}
\end{equation}
We now consider the O.P.E. of $T$ with the deformation operators 
${\cal C}$, ${\cal D}$ (\ref{recoilops}). 

The computation of the time-dependent parts of $T$, 
$2T_t \equiv -(\partial_z t)^2$, and of the space-dependent ones,   
$2T_X \equiv a^2(t) (\partial_z X^i)^2$, may be 
carried out separately~\cite{grav}.
The basic ingredients are Eq.(\ref{propagator}), 
together with the standard representation (\ref{regulator}) 
of the Heaviside operator, as well as the expansion 
\begin{equation}
(\partial X^j(z))^2 \odot \partial_n X^i(w) \sim 
G^{ii}\frac{1}{(z-w)^2}\partial_n X^i \sim \frac{a_0^{-2}e^{-2Ht}}{(z-w)^2}
\partial_n X^i, \qquad ({\rm no}~~{\rm sum}~~{\rm over}~~i)
\label{spatialparts}
\end{equation}
The analysis is straightforward, and yields:
\begin{eqnarray}
&& 2T_t(z) \odot {\cal C}(z') \sim 
\frac{1}{|z - z'|^2}\left(-\frac{\epsilon^2}{2}\right)
{\cal C}(z')~, \nonumber \\
&& 2T_X(z) \odot {\cal C}(z') = {\tilde c}_i 
\Theta_\epsilon (t - 2H{\rm ln}|z - z'|)
\frac{\partial_n X^i}{|z - z'|^2} = {\tilde c}_i \Theta_\epsilon (t) 
\frac{\partial_n X^i}{|z - z'|^{2-2H\epsilon}}~, \nonumber \\
&& 2T_t(z) \odot {\cal D}(z') \sim 
\frac{1}{|z - z'|^2}\left(-\frac{\epsilon^2}{2}\right)
{\cal D}(z')~, \nonumber \\
&& 2T_X(z) \odot {\cal D}(z') = c_{0,i} 
\Theta_\epsilon (t - 2H{\rm ln}|z - z'|)
\frac{\partial_n X^i}{|z - z'|^{2 + 4H^2}} e^{2Ht (z)}= \nonumber \\
&& c_{0,i} \Theta_\epsilon (t)
\frac{\partial_n X^i}{|z - z'|^{2 + 4H^2 - 2H\epsilon}}e^{2Ht (z)}~, 
\label{opes}
\end{eqnarray}
where $\epsilon \to 0^+$, and we 
used Taylor expansion in the arguments of the Heaviside operators, 
and the property $d\Theta_\epsilon (t) / dt = -\epsilon \,\Theta_\epsilon $. 

Thus we arrive at the following result to leading order in 
world-sheet divergences as $z \to z'$: 
\begin{eqnarray}
&&2T(z) \odot {\cal C}(z') \sim \frac{1}{|z - z'|^2}{\cal C}(z)~, \nonumber \\ 
&&2T(z) \odot {\cal D}(z') \sim \frac{1}{|z - z'|^{2 + 4H^2}}{\cal D}(z)~,
\label{tope}
\end{eqnarray}
Notice that the {\it Minkowski signature} 
of the target space-time is {\it crucial}
to the effect of yielding positive anomalous scaling dimension 
for the ${\it D}$ operator, proportional 
to $H^2$ (notice that the operator ${\cal C}$ is conformal in the limit 
$\epsilon \to 0^+$). Due to the non-renormalizability of the 
world-sheet stress tensor in two dimensions, $T$, one infers from 
(\ref{tope}) that the operator ${\cal D}$ has anomalous scaling dimension
$4H^2$ in string 
units ($M_s=1$), i.e. $4H^2/M_s^2$ in arbitrary units. 
We therefore observe that, in the regime of parameters (\ref{relconstr})
we are working with, 
such terms doninate over 
space-time curvature 
effects that are neglected in (\ref{propagator}).

For consistency, one should verify 
the above result 
by computing the O.P.E.'s 
of the operators ${\cal C},~{\cal D}$
with themselves.  
The computation confirms the conformal nature of 
${\cal C}$, while for ${\cal D}$ one obtains:
\begin{equation}\label{finalope}
{\cal D}(z) \odot {\cal D}(z') \sim \frac{1}{|z - z'|^{2 + 4H^2}}{\cal D}(z')
\end{equation}
implying its anomalous dimension $4H^2$, and hence its world-sheet 
renormalization group relevance. 
The appearance of the anomalous dimension 
implies that the recoil process has spoiled
the conformal  invariance of the $\sigma$-model, 
which now needs Liouville dressing~\cite{ddk} to restore it. 

It must be noticed at this stage that above we have 
considered the de Sitter space-time as a valid stringy 
$\sigma$-model background,
and concentrated on the effects of recoil, assuming the latter 
to be the only source of (boundary) world-sheet violation of
conformal invariance. In the Appendix of the current article 
we discuss a model in which the de Sitter space-time itself is
viewed as a solution of generalized conformal invariance conditions
of a non-critical Liouville string~\cite{emninfl}, expressing the 
restoration of the conformal invariance by the Liouville mode~\cite{ddk}.
In this more complete picture the recoil deformation constitutes 
one of the possible non-critical 
deformations of the specific Liouville $\sigma$-model under consideration, 
whose deficit $Q^2$ causes 
inflation in the way explained in the 
Appendix. 
The formalism described above remains intact for this (mathematically 
complete) case. As will be discussed in the next section,
this is so because  
the effects of the recoil on the space-time are such that they 
may be removed 
by means of an appropriate coordinate transformation; thus, the 
space-time emerging after the recoil (written in 
the new coordinates) retains its de-Sitter form.

\section{Liouville Dressing and Recoil-Induced Space-Time Distortions}

As discussed in \cite{grav}, there are two equivalent ways of 
performing the Liouville dressing: one is to dress the boundary operator
${\cal D}$ as it stands, and restore conformal invariance on the 
boundary world-sheet theory, and the other (which we shall follow here) 
is to write first the operator ${\cal D}$ as a total world-sheet derivative
bulk operator (using Stoke's theorem), 
and then dress the bulk operator. In particular, in this latter case we have: 
\begin{eqnarray} \label{bulk} 
&& V_{L, {\rm bulk}} = \int _{\Sigma} e^{\alpha_i \varphi} 
\partial_\alpha \left(y_i(t)\partial^\alpha X^i \right)= 
c_{0,i}\int_{\Sigma} e^{\alpha_i \varphi} 
\partial_\alpha \left(\Theta_\epsilon (t) e^{2Ht}\partial^\alpha X^i \right)~,
\nonumber \\
&& c_{0,i} = a_0^2\frac{u^i}{2H}~, \qquad 
\alpha_i = -\frac{Q}{2} + \sqrt{\frac{Q^2}{2} + (2-\Delta_i)}
\end{eqnarray} 
where $\varphi (z)$ is the world-sheet Liouville field,   
$\Delta_i$ is the conformal dimension of the bulk operator,
with $2 - \Delta_i$ its anomalous dimension, 
and $\alpha_i $ is the 
so-called gravitational anomalous dimension.
The central charge deficit $Q^2$ is the one responsible 
for the initial inflationary phase, and 
may be estimated as follows: one may consider various scenaria 
for departure from criticality, as necessary for inflation 
in stringy $\sigma$-models~\cite{emninfl}; for example, 
in \cite{grav} this was  
due to ``catastrophic'' cosmic events, such as the collision
of two brane worlds. In such a scenario, which is 
discussed briefly in the Appendix, it is possible to 
obtain an initial {\it supercritical } central charge deficit 
(and hence a time-like Liouville mode~\cite{aben})
of order 
\begin{equation}\label{centraldeficit} 
Q^2 = 9 H^2 > 0~,
\end{equation}
where the Hubble parameter $H$ can be fixed in terms of other
parameters in the theory; for instance, in the 
specific model of \cite{grav} on colliding branes,
$Q$ (and thus $H$), is found to be proportional to the 
square of the relative velocity of the colliding branes, 
$Q \propto u^2$ during the inflationary era. 
As we show in the Appendix, in a phase of constant $Q$, one 
obtains an inflationary de-Sitter type 
Universe~\footnote{We note in passing that cosmically 
catastrophic non-critical
string scenaria, as the one of ref.\cite{grav2}, 
allow for a {\it relaxing to zero} deficit $Q^2(t)$
in such a way that, although  during the inflationary era $Q^2 $ is 
(for all practical purposes)
constant, as in (\ref{centraldeficit}), eventually $Q^2$ decreases 
with time so that, 
at the present era,
one obtains compatibility with an accelerating Universe phase,
as suggested by a variety of recent astrophysical observations. 
As already mentioned, 
such relaxation (quintessence-like) scenaria~\cite{grav,papant} 
have the advantage of properly definable  
asymptotic states (as $t \to \infty$) and string scattering $S$-matrix.
Other scenaria for inducing de Sitter Universes in string theory
may be envisaged, where the inflation space-time is obtained 
as a result of string loops (dilaton tadpoles)~\cite{fischler},
but in such models a string $S$-matrix cannot be properly defined.}.

The specific normalization in (\ref{centraldeficit}) 
is imposed because, as discussed below and in the Appendix, 
(i) one may identify the time $t$ with a 
Liouville mode $-\varphi$ 
of the {\it supercritical } $\sigma$-model: the minus sign  
is justified below both mathematically, 
due to properties of the Liouville mode, and 
physically by the requirement of a relaxation 
to zero of the recoil-distorted space-time deformation. 
(ii), under this identification, the Liouville 
equation for the modes/$\sigma$-model background couplings~\cite{emn}:
\begin{equation}
{\ddot g}^i + Q{\dot g}^i 
= -\beta^i (g) = -{\cal G}^{ij} \partial C[g]/\partial g^j~,
\label{liouveq}
\end{equation}
where the dot denotes derivative with respect to the Liouville
world-sheet zero mode $\varphi$, and 
${\cal G}^{ij}$ is an inverse Zamolodchikov metric in 
string theory $\{ g^i \}$ space, 
when applied to scalar (e.g. inflaton-like) string modes would yield standard
field equations for scalar fields in de Sitter (inflationary)   
space-times provided the normalization 
(\ref{centraldeficit}) is valid. This would imply 
a ''Hubble'' parameter $Q=3H$ (notice that 
the gradient flow property of the 
$\beta$ functions makes the analogy with the inflationary case even more 
profound, where the running central charge $C[g]$~\cite{zam} plays the 
r\^ole of the inflaton potential in conventional inflationary field theory). 

As discussed in the Appendix, there are consistent solutions 
of the generalized conformal invariance conditions with the above described 
desired properties~\footnote{At this stage we remark 
that, since in general the non-critical string scenario 
produces inflationary phases dynamically, due to the 
presence of the Liouville mode~\cite{emninfl,papant,grav2},
without introducing extra inflaton fields, 
there is no need for the requirement that the equations (\ref{liouveq})
for scalar string modes be formally identical to those of an  
inflaton field in conventional cosmology. 
The precise proportionality 
coefficient between $Q^2$ and $H^2$ 
is actually physically unimportant 
in our framework, as long as it is non-zero. 
Indeed, such a constant will only affect 
the precise expression for the Liouville anomalous dimension $\alpha_D$ 
(\ref{gravdim}) (c.f. below) $\alpha_D = \gamma H$, where $\gamma \ne 2$ 
for any $Q^2 \ne 0$, and $\gamma = 2$ if and only if $Q=0$ (no
central-charge deficit). Such constants $\gamma \ne 2$ do not affect qualitatively
our results and can be absorbed in normalization factors. 
Thus, the above normalization (\ref{centraldeficit}) 
between $Q^2$ and $H^2$ can actually change, and is model dependent.
What is model independent (within the class of models that yield 
string inflation), though, is that the deficit $Q^2$ is always proportional 
to the square of the Hubble 
parameter $H$ appearing in the exponential of 
the Universe scale factor $a(t)=a_0 e^{Ht}$.
However, it is comforting (from a phenomenological viewpoint at least) 
that a normalization (\ref{centraldeficit}), which  
is in agreement with conventional inflationary cosmology, as far as 
the inflaton field is concerned, can be found consistently 
in our approach.}.

We now remark that 
the relations (\ref{liouveq}) replace the 
conformal invariance conditions $\beta^i = 0$ of the critical string, and  
serve in expressing the necessary conditions for the 
restoration of conformal invariance by the Liouville 
mode~\cite{ddk}. In fact, upon interpreting the latter as an extra target
dimension, the conditions (\ref{liouveq}) may also be viewed as 
conformal invariance conditions of a (D+1)-target-space dimensional
{\it critical} $\sigma$-model (D is the target dimension of the 
non-critical $\sigma$-model before Liouville dressing). 
In most Liouville approaches one treats the Liouville mode $\varphi$ 
and time $t$ as independent coordinates. 
In our approach \cite{emn,papant,grav2}, however, we take one step further,
and provide dynamical arguments which imply the 
restriction in this extended (D+1)-dimensional space time 
of lying on a hypersurface determined by 
the identification $\varphi = -t$. 
This means that, as the time flows 
in our Universe, we are forced to lie on this D-dimensional subspace
of the (D+1)-dimensional Liouville extended 
space time~\footnote{For instance, in the work of \cite{grav2}, 
involving brane-world collisions as a source of 
departure from criticality,
this restriction
was imposed because the potential of massive particles,
in an effective field theory context, 
was found to be proportional to ${\rm cosh}(t + \varphi)$,
which is thus minimized at $\varphi = -t$.
In fact,
such an opposite flow of the Liouville mode as compared to that of 
target time
may be given a deeper mathematical interpretation.  
It may be 
viewed as a consequence of a specific treatment 
of the area constraint 
in non-critical (Liouville) $\sigma$-models~\cite{kogan1,emn},
which involves
the evaluation of the Liouville mode 
path integral via an appropriate steepest-descent 
contour. 
In this way one obtains 
a `breathing' world-sheet 
evolution, in which the world-sheet
area starts from a very large value (infrared cutoff), shrinks to  
a very small one (ultraviolet cutoff), and then inflates again 
towards very large values (infrared
cutoff).
Such a situation may then be interpreted 
as 
a world-sheet `bounce' at the infrared,
implying, following the reasoning of 
ref.~\cite{coleman}, that the physical 
flow of target time is 
opposite to that of the world-sheet scale (Liouville zero mode).}.

{}From (\ref{finalope}), (\ref{centraldeficit}),  
(\ref{bulk}), and using that  $2 - \Delta_i = 4 H^2 $
one has:
\begin{equation} 
\alpha _{{\cal D}} = H 
\label{gravdim}
\end{equation}
Then, after 
performing 
a world-sheet integration by parts in (\ref{bulk}), 
it is straightforward to obtain a non-diagonal deformation in the Liouville-extended
$(\varphi, t, X^i) $ target space metric:
\begin{equation}\label{liouvillmetric} 
G_{\varphi i} = -Hc_{0,i} \Theta_\epsilon (t) e^{H(\varphi + t)}~,~
c_{0,i} = a_0^2\frac{u^i}{2H}.
\end{equation}
The resulting extended space-time distortion then is 
given by the following line element (for times $t \gg 0$, i.e. long 
after the scattering event):
\begin{equation}\label{recoilmetricfl} 
ds^2 = -(d\varphi) ^2 - (dt)^2 + e^{2Ht}\left( a_0^2(dX^i)^2 + 2Hc_{0,i} 
e^{H\varphi}d\varphi dX^i\right)~,~ c_{0,i} = a_0^2\frac{u^i}{2H}.
\end{equation}
Upon the identification $\varphi = -t$ one then 
obtains an expression for the distorted space-time 
due to D-particle recoil, for $t > 0$:
\begin{equation}
ds^2 = -2(dt) ^2 + e^{2Ht}\left( a_0^2(dX^i)^2 + 2Hc_{0,i} 
e^{-Ht}d\varphi dX^i\right)
\label{distorted}
\end{equation}
This result is consistent with
our perturbative (world-sheet) treatment of inflation, which we consider to  
start at $t =0$, in the sense 
that as $Ht $ becomes larger
the distortion effects become considerably smaller 
(exponentially suppressed). Thus, the distortion effects  
will be washed out completely by the exponential
expansion, as expected. However, as we shall discuss in the next section, 
global effects due the presence
of D-particles will leave their
trace through small angular modifications in the particle production,

To study such effects, we first observe that the distorted 
space-time (\ref{distorted}) can be cast in a conventional 
de Sitter inflationary form
\begin{equation}
ds^{2} = -(dt')^{2} + e^{\sqrt{2}Ht'}(a_0^2d{\tilde X})^{2}
\label{dds}
\end{equation}
upon changing coordinates $(t, X^{i}) \to (t', {\tilde X}^{i})$:
\begin{eqnarray}
&& t' = \sqrt{2}t - \frac{H}{2}|c_{0,i}|^{2}e^{-2Ht} = 
\sqrt{2}t - \frac{a_0^4}{8H}|u^i|^{2}\left(e^{-2Ht} -1 \right)~, \qquad \nonumber\\
&& {\tilde X}^i = X^{i} + a_0^2 \frac{u^i}{2H} \left(e^{-Ht} - 1 \right)
\label{newcoordinates}
\end{eqnarray} 
where smoothness of the limit $H \to 0^+$ has been 
guaranteed by imposing  appropriate boundary conditions. 
Thus, one arrives at the following two space-time geometries,
before and after the D-particle collision:
\begin{eqnarray}
&&(ds_{\rm IN})^2 = -2(dt)^2 + a_0^2 (dX^i)^2~, \nonumber \\ 
&&(ds_{{\rm t} > 0})^2 = -(dt')^2 + a(t')^2 (d{\tilde X}^i)^2~,
\label{bogolubov}
\end{eqnarray}
where then ${\rm IN}$ vacuum occurs at $t \to -\infty$, while 
$(t', {\tilde X}^i)$ occur for $t \ge 0$ and are 
given in (\ref{newcoordinates}). 
In the next section we will use the above modification in the two geometries
in order to compute the corresponding particle production:
one will seek an 
appropriate (smooth) interpolating function $a(t)$ between 
the IN flat space-time (which is assumed to occur 
from $t=-\infty$ till $t= 0$) and the ${\rm OUT}$ space-time,
which should include inflation in the regime $0 <  Ht \ll 1$ 
consistent with our $\sigma$-model considerations (we stress again that 
the moment of impact of the string with the D-particle is at $t=0$). 
Before closing this section we 
would like to mention 
that the effects of D-brane recoil in flat background space-time
have been considered in \cite{winst}, and a non-thermal,
non-isotropic particle production has also been demonstrated
in that case.

\section{Particle Production in D-particle-recoil-Distorted 
Inflationary Universe}

For the benefit of the reader we would like first to 
outline the general method for computing particle production
in Robertson-Walker space-times~\cite{chung}. 

Consider for definiteness a {\it massive} scalar field $X$,
of mass $M_X$, 
on a curved Robertson-Walker (RW) background, in the conformal 
frame $(\eta, X^i )$ which is defined as~\cite{birrell}:
\begin{equation}
ds^2 = a^2(\eta ) \left( -(d\eta)^2 + (dX^i)^2 \right)
\label{conformal}
\end{equation}
This frame is related to the RW (cosmological comoving) 
frame $(t, X^i)$ used in the previous sections
by: 
\begin{equation}
d\eta = \frac{dt}{a(t)}
\label{comovingtime}
\end{equation}
For the inflationary case of $a(t)= a_0e^{Ht}$, we have that the corresponding conformal time 
is given by 
\begin{equation}
\eta = - \frac{1}{H} \bigg(\frac{1}{a_0}e^{-Ht}-1\bigg)
\label{etatime}
\end{equation}
where the integration constant has been chosen such that $\eta\to t$ as $H\to 0$. 

One may expand the field $X$ in Fourier modes $h_k$, defined through the 
Klein-Gordon equation that $X$ obeys   
in such metric backgrounds. In particular, in the comoving 
frame $(t, X^i)$: 
\begin{equation}
X = \int \frac{d^3 k}{(2\pi)^{3/2}a(t)}\left[a_k e^{i {\vec k} \cdot {\vec X }}h_k (t) + a_k^\dagger e^{-i {\vec k} \cdot {\vec X }}h_k^* (t) \right]
\label{modes}
\end{equation}
where the vectors denote spatial components of four vectors.
In the conformal time frame one arrives at the following 
equation (stemming from the Klein Gordon equation): 
\begin{equation}
h_k^{''}(\eta ) + \omega_k^2(\eta) h_k(\eta)  = 0
\label{modeeq}
\end{equation}
where 
the prime denotes 
differentiation with respect to $\eta$, and 
the energy (positive frequency) is given by 
$\omega_k = +\sqrt{k^2 + M^2 C(\eta)}$,
where $C(\eta) = a^2\left( 1 + {\rm interactions}\right)$
is an effective scale factor~\cite{chung} including the 
effects of interaction of the $X$ field with the inflaton
background {\it etc}.  

After casting $h_k$ in the form 
\begin{equation}
h_k = \frac{\alpha_k}{\sqrt{2\omega }}e^{-i\int^\eta  \omega_k d\eta ' }
+ \frac{\beta_k}{\sqrt{2\omega }}e^{i\int^\eta  \omega_k d\eta ' }
\end{equation}
one obtains a system of differential (first order) 
equations for $\alpha_k, \beta_k$, which are Bogolubov~\cite{birrell}
coefficients connecting the two vacua (IN and OUT) 
satisfying $\alpha_k ^2 - \beta_k ^2 = 1$:
\begin{eqnarray}
\alpha_k' &=& \frac{\omega_k'}{2\omega_k} \,\beta_k
{\rm exp}\left(-2i\int _{\eta_p}^{\eta}\omega _k (\eta ') d\eta '\right)
\nonumber\\
\beta_k' &=&  \frac{\omega_k'}{2\omega_k} \,\alpha_k
{\rm exp}\left(-2i\int _{\eta_p}^{\eta}\omega _k (\eta ') d\eta '\right)
\end{eqnarray}
 
As mentioned earlier, for a consistent interpretation of $|\beta_k|^2$
as  a particle density,  
it is {\it imperative}~\cite{schwarz} 
that, asymptotically in the (conformal ) time 
$\eta \to \pm \infty$, the space-time be {\it flat }.
This is so because, it is only in this 
case that the modes used in the expansion of the field $X$ 
can be put in a suitable relation to the known modes of Minkowski space-time.
This will be assumed in what follows, which will necessitate 
viewing the inflationary epoch of the Universe only as a specific 
era, occurring in a space-time which interpolates smoothly 
(in $\eta$ time) between two asymptotically flat regions.
{}From our world-sheet  (Liouville $\sigma$-model) 
point of view this means that 
the inflationary (non-conformal) space-time background 
connects (in the sense of a running along a world-sheet renormalization
group (RG) trajectory) two asymptotic
fixed points of the RG flow (flat space-time regions). 
Upon our identification of the RG flow with the actual target-space 
time flow of the $\sigma$-model~\cite{emn}, the RG evolution 
becomes in this way a cosmological evolution, implying 
graceful exit from the de Sitter (inflationary) era, 
and thus allowing for the definition of an S-matrix.

In our considerations in this article we shall use some simplified
interpolating metrics appearing in the literature, 
which may not be exact solutions of (Liouville) string theory. 
Their main advantage is that they can be 
treated analytically. For more detailed studies within the context
of realistic (Liouville) string models, allowing for a graceful exit
from inflation, see \cite{papant}; such metrics, however, can only be studied numerically.

Assuming the above, one can apply WKB (essentially derivative expansion) 
methods~\cite{chung,biswas} 
to approximate the Bogolubov coefficient $\beta_k$ as~\cite{chung}:
\begin{equation}
\beta_k \simeq \int d\eta \frac{\omega_k'}{2\omega_k}{\rm exp}
\left(-2i\int _{\eta_p}^{\eta}\omega _k (\eta ') d\eta '\right)
\label{bobetax}
\end{equation}
to leading order, with the boundary conditions $\alpha_k (\eta_p) =1,~\beta_k (\eta_p) =0$. 
In deriving the approximate expression of Eq.(\ref{bobetax})
one has to make sure that  
the conditions for the validity of the WKB
approximation are satisfied at all times; this in turn leads 
to restrictions, which are discussed in detail in ref. \cite{chung},
where we refer the interested reader. 

To evaluate the integral approximately, one can apply a steepest-descent 
method, by complexifying the 
time $\eta$, and then approximating the value of the integral by a sum of 
values of the integrand at the branch-cut points of a steepest-descent
contour, which can be appropriately constructed.
The starting points ${\tilde \eta}_j$ 
of the branch-cut  in the complex $\eta$ plane
are defined as:
\begin{equation}
\omega_k ({\tilde \eta}_j ) = 0 = k^2 + M^2 C(\eta_j) 
\label{branchcuts}
\end{equation}
There are two approximations involved~\cite{chung}: one is the above-mentioned 
approximation of the integral over $\eta$ in (\ref{bobetax}) 
as a sum over branch-cut contributions, and the second is a derivative
expansion of the integral over $\eta '$ in the exponent near each branch-cut
point. To lowest order in derivatives one may write:
\begin{equation}
\int _{\eta_p}^{\eta} \omega_k (\eta ') d\eta ' \simeq 
\int _{\eta_p}^{{\tilde \eta}_j} \omega_k (\eta ') d\eta ' 
+ \frac{2M_X}{3}\sqrt{C'({\tilde \eta}_j)}\delta ^{3/2} + \dots 
\end{equation}
with $\delta \equiv \eta - {\tilde \eta}_j $.

\begin{figure}[t]
\begin{center}
  \epsfig{file=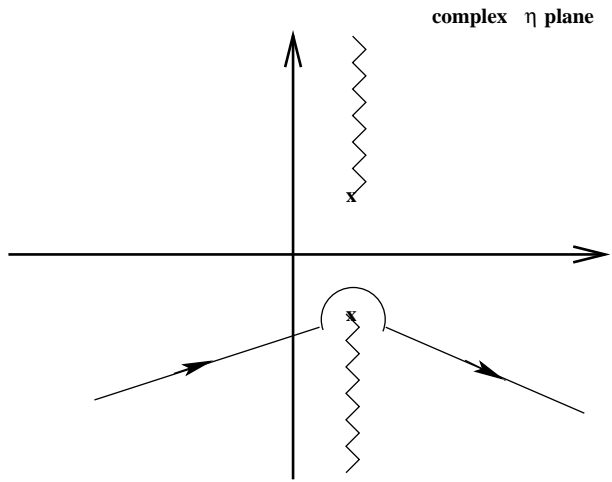,angle=0,width=0.5\linewidth} 
\end{center}
\caption{\it Steepest descent contour and branch cut structure
for the integral (\ref{bobetax}).}
\label{sdcontour}
\end{figure}

In this way one may {\it approximate}  the coefficient $\beta_k$ (\ref{bobetax})
as a sum over branch cut points:
\begin{equation}
\beta _k \simeq \sum _{j} {\cal U}_j~, \qquad 
{\cal U}_j \equiv 
{\rm exp}\left(-2i\int_{\eta_p}^{{\tilde \eta}_j}\omega _k (\eta ') 
d\eta ' \right)
\int_{{\cal C}_j}\frac{d\delta }{\delta }{\rm exp}
\left(-\frac{4i}{3}M_X\sqrt{C'({\tilde \eta}_j)}\delta ^{3/2} + \dots\right) 
\label{approxbc}
\end{equation}
where $C_j$ is a steepest descent contour. 
In the examples considered in \cite{chung}, and also here, 
the appropriate contour is the one that encompasses the 
branch cuts that lie in the lower half plane (see fig.  
\ref{sdcontour}). 
This is associated with the fact that 
one chooses the sign $\sqrt{C'(\eta)} > 0$, due to 
the positive frequency requirement; 
the branch cuts have been chosen to go towards
$\pm \infty$, and a steepest descent 
contour {\it must} have an ``incoming'' 
and an ``outgoing'' direction~\cite{chung}. 

An important approximation in the 
WKB analysis of \cite{chung}, 
which as we shall see remains valid in our case here, is that 
{\it only one} pole dominates the integral (\ref{approxbc}).
Let the value of ${\tilde \eta}_j$ in this dominant 
pole be split in real ($r$) and imaginary ($\mu$) parts:
$\eta = r + i \mu$.  
This, 
then, yields 
\begin{equation}
|\beta _k |^2 \simeq {\rm exp}\left(-4\bigg[\frac{(k/a_{\rm eff}(r))^2}{M_X
\sqrt{H^2_{\rm eff}(r) + R_{\rm eff}(r)/6}} + 
\frac{M_X}{\sqrt{H^2_{\rm eff}(r) + R_{\rm eff}(r)/6}}\bigg]\right)
\label{inflationbog}
\end{equation}
up to irrelevant proportionality constants of order ${\cal O}(1)$, 
with $a_{\rm eff} = \sqrt{C}$, $H_ {\rm eff}$ the effective Hubble parameter,
and $R_{\rm eff}(r)$ the effective scalar curvature (which in our case 
may be neglected). 
The density of the $X$ particles~\cite{chung}, 
created at time $t$ in comoving frame, is given by
\begin{equation}
n_X(t) = \int_0^{\infty}\frac{dk}{2\pi^2 a^3(t)} k^2 |\beta _k |^2 
\label{creation}
\end{equation}

If we now assume, as in \cite{chung}, 
that most of the support of the integral comes from $(k/M_X)^2 \ll 1$, 
where the $k$-dependence of $r$ may be neglected, we arrive at     
\begin{equation}
n_X \simeq \frac{a^2_{\rm eff}(r)}{8\pi ^{3/2}a^3(t)}
{\rm exp}\left(\frac{-4M_X}{\sqrt{H^2_{\rm eff}(r) + R_{\rm eff}(r)/6}}\right)
\times 
\left(\frac{M_X}{4}\sqrt{H^2_{\rm eff}(r) + R_{\rm eff}(r)/6}\right)^{3/2}
\label{density}
\end{equation}

The above considerations pertain to the standard RW cosmology, including a de Sitter phase, provided one connects the latter smoothly to 
asymptotically flat regions~\cite{schwarz}. 

After these general considerations, we now turn 
to estimate the effects on the particle production
due to the distortion in the underlying metric induced  by the initial D-brane recoil.
To accomplish that we must first determine the modifications produced to the 
corresponding Bogolubov coefficient $\beta _k$, due to 
the presence of the terms proportional to $u^i$ in the coordinate transformation of
(\ref{newcoordinates}). 

Keeping only terms linear in $u^i$, 
and introducing the conformal time $\eta$, (\ref{newcoordinates}) becomes
\begin{eqnarray}
&& t' = \sqrt{2}t \qquad \nonumber\\
&& {\tilde X}^i = X^{i} - \frac{1}{2}a_0^2  u^i \eta 
\label{newcoordinates2}
\end{eqnarray} 

Then, under the influence of the recoil, 
the corresponding 
Bogolubov coefficient, to be denoted by $\beta _k^{(u)}$, reads
\begin{equation}
\beta_k^{(u)} \simeq \int d\eta \frac{\omega_k'}{2\omega_k}
{\rm exp}\left(-2i\int _{\eta_p}^{\eta}\omega _k (\eta ') d\eta '\right)
{\rm exp}\left[-  \frac{i}{2} \, a_0^2  (u_i k^i) \eta  \right]
\label{bobu}
\end{equation}
Under the same WKB assumptions, the branch cut contributions ${\cal U}_j$ 
of (\ref{approxbc}) are now modified to  
\begin{eqnarray}
{\cal U}_j^{(u)} &=& 
{\rm exp}\left(-2i\int_{\eta_p}^{{\tilde \eta}_j}\omega _k (\eta ') 
d\eta '   - \frac{i}{2} a_0^2 (u_i k^i) {\tilde \eta}_j \right) \nonumber\\
&\times & \int_{{\cal C}_j}\frac{d\delta }{\delta }{\rm exp}
\left(-\frac{4i}{3}M_X\sqrt{C'({\tilde \eta}_j)}\delta^{3/2}  
- \frac{i}{2} a_0^2 (u_i k^i) \delta 
\dots\right)
 \label{UJ}
\end{eqnarray}
Notice at this point that the additional contribution due to the recoil 
does not alter the branch-cut structure of our problem, which is still determined 
by  (\ref{branchcuts}) and by the sign of $\sqrt{C'(\eta)}$; the latter 
may still be chosen positive as in \cite{chung}. 
It turns out that, under the assumptions of \cite{chung}, 
the additional term in the integral over $\delta$
does not change the result of \cite{chung}.  
This is so because the value is determined solely by the 
angular integration, since the 
steepest descent contour 
evades the beginning of the branch point
in such a way that~\cite{chung} $\delta = \epsilon e^{i\theta}$, 
$\epsilon \to 0^+$ and $\theta  
\in (-\frac{2\pi}{3}, \frac{4\pi}{3} )$, 
along the semicircle around it (c.f. figure \ref{sdcontour}). 
On the other hand,  
the imaginary part of ${\tilde \eta}_j$ in the first exponential 
gives a non-vanishing 
contribution, whereas  its real part contributes only a phase; the latter
does not contribute in the physically relevant $|\beta _k |^2$. Thus, 
 setting ${\tilde \eta}_j = \tilde{r}_j + i \tilde{\mu}_j$, 
and assuming again one dominant pole we have  
\begin{equation}
|\beta _k^{(u)}|^2 = |\beta _k|^2  
{\rm exp}\left[ a_0^2 (u_i k^i) \tilde{\mu} \right]
\label{bogu}
\end{equation}
In the case that $C(\eta)$ is purely of inflationary origin, i.e. 
$C(\eta) = (H \eta  -1)^{-2}$, the solution of (\ref{branchcuts}) gives 
$\tilde \eta^{\pm} = H^{-1} (1 \pm i M/|k|)$; choosing the lower half plane,
we have that $\tilde{\mu} = - H^{-1} (M/|k|)$. Then, setting 
$u_i k^i = |u||k|\cos\theta$, where $\theta$ is the angle between the 
velocity of the original D-particle and the emitted (produced) particle, 
(\ref{bogu}) yields
a $k$-independent correction term to $|\beta _k^{(u)}|^2$ of the form 
\begin{equation}
|\beta _k^{(u)}|^2 = |\beta _k|^2 {\rm exp}\left( - a_0^2 |u| H^{-1} M  \cos\theta\right)
\end{equation}
which in turn modifies the particle number production from (\ref{creation}) to 
\begin{equation}
n_X^{(u)} (t) = n_X(t) \, {\rm exp}\left( -  a_0^2 |u| H^{-1} M  \cos\theta\right)
\label{creu}
\end{equation}
Notice that in the absence of inflation, i.e. if $H\to 0$, 
the additional effect vanishes exponentially.
The above result seems to persist once one takes into account
interpolating metrics, with asymptotically flat regions, as 
required for the proper interpretation of the square of the Bogolubov
coefficient $|\beta_k|^2$ as particle number. 
For example, consider the RW metric with effective 
scale factor~\cite{gubser}
\begin{equation}
C(\eta) = A + B\frac{H\eta - 1}{\sqrt{(H\eta - 1)^2 + 1}}
\label{gubser}
\end{equation}
which, for appropriate choice of
the parameters ($A=-B >0$) contains 
a region of standard inflationary evolution, i.e. 
$1/(H\eta - 1)^2$, for $(H\eta -1)^2 \gg 1$, 
and interpolates between asymptotically flat metrics. 
To ensure that the inflationary phase of this interpolating metric 
coincides with our time regime $0 \ll Ht \ll 1$ 
for the validity of the $\sigma$-model perturbation theory, 
one must impose the condition $a_0 \ll 1$ in the relation (\ref{etatime})  
connecting the conformal time $\eta$ with the Robertson-Walker time $t$. 
In practical terms,
$a_0$ can be taken to be two to three orders 
of magnitude less than unity (in Planck units), yielding  
$|H\eta - 1| \sim 10^3$ during inflation, for times $0 < Ht \ll 1$. 
It goes without saying that the metric (\ref{gubser}) is used
only for illustrative purposes, and is not directly related to any
realistic string model. At present, within the context of the non-critical
string approach to inflation, interpolating metrics including 
inflation at an early stage of their evolution have only been 
obtained numerically for realistic (Liouville) string theories~\cite{papant}.

The branch cut structure of the metric (\ref{gubser}) leads again to 
two branch cuts, with the (relevant) one lying in the lower half plane
being given by: ${\tilde \eta} = \frac{1}{H}-i\frac{\sqrt{k^2 +A M^2}}{H |k|}$.
Then, the corresponding $|\beta _k^{(u)}|^2$ displays a $k$-dependence, 
\begin{equation}
|\beta _k^{(u)}|^2 = |\beta _k|^2 {\rm exp}\left( - a_0^2 |u| H^{-1} \sqrt{k^2 + A M^2 }  
\cos\theta\right)
\label{spectrumfinal}
\end{equation}
If $k^2$ can be neglected next to $M^2$, which is what one usually assumes, then the 
results with and without interpolating metrics coincide. 

Before turning into the discussion of the possible physical implications of the above results,
we would like to comment on an additional point.  Returning to (\ref{bulk}), we see that 
in the case of a critical Liouville string, i.e for $Q^2=0$, the corresponding 
$\alpha _{{\cal D}}$ would be  $\alpha _{{\cal D}} = 2H$ instead of the one given in 
(\ref{gravdim}). In that case, the corresponding ${\tilde X}^i$ in 
(\ref{newcoordinates})
would then become
\begin{eqnarray}
{\tilde X}^i &=& X^{i} + a_0^2 \frac{u^i}{2H} \left(e^{-2Ht} - 1 \right) \nonumber\\ 
&=& X^{i} + a_0^2 u^i \eta \bigg( \frac{H \eta}{2} - 1\bigg)
\end{eqnarray}
Then, the contribution due to the recoil in the expressions of (\ref{UJ})
would be 
\begin{equation}
{\cal U}_j^{(u)} = 
{\rm exp}\left(-2i\int_{\eta_p}^{{\tilde \eta}_j}\omega _k (\eta ') 
d\eta '  +i L_{{\tilde \eta}_j}^{(u)}\right)
\int_{{\cal C}_j}\frac{d\delta }{\delta }{\rm exp}
\left(-\frac{4i}{3}M_X\sqrt{C'({\tilde \eta}_j)}\delta^{3/2}  
+i L_{\delta}^{(u)}
\dots\right) 
\end{equation}
where the $\delta$-dependent factor  $L_{\delta}^{(u)}$ and the 
$\delta$-independent factor  $L_{{\tilde \eta}_j}^{(u)}$ are given by
\begin{eqnarray}
L_{\delta}^{(u)} &=& a_0^2 (u_i k^i)
\bigg[\frac{H}{2}\, \delta^2 + (H {\tilde \eta}_j  -1)\delta \bigg]
\nonumber\\
L_{{\tilde \eta}_j}^{(u)} &=& a_0^2 (u_i k^i)
{\tilde \eta}_j \bigg( \frac{H {\tilde \eta}_j}{2} - 1\bigg)
\end{eqnarray}
Again, only the imaginary parts of $L_{{\tilde \eta}_j}^{(u)}$, to be denoted by 
$\Im m L_{{\tilde \eta}_j}^{(u)}$, 
give a non-vanishing 
contribution, whereas  its real part contributes only a phase, i.e. 
\begin{equation}
{\cal U}_j^{(u)} = {\cal U}_j  
{\rm exp}\left( - \Im m L_{{\tilde \eta}_j}^{(u)} \right)
\end{equation}
and after setting again ${\tilde \eta}_j = \tilde{r}_j + i \tilde{\mu}_j$ we have 
(assuming again one dominant pole)
\begin{equation}
{\cal U}_j^{(u)} = {\cal U}_j  
{\rm exp}\left[- a_0^2 (u_i k^i) \tilde{\mu} (H \tilde{r} -1) \right]
\end{equation}
Then for the purely inflationary part of the 
metric, $\tilde \eta^{\pm} = H^{-1} (1 \pm i M/|k|)$ and therefore $H \tilde{r} -1 = 0$; thus, 
when the string is critical, the bulk of the effect from the recoil vanishes, as it should. 
This result persists as long as $\tilde{r} = H^{-1}$, as is the case of the interpolating 
metric of  (\ref{gubser}).
We consider this as a non-trivial 
self-consistency check of both our method and of the integration techniques developed 
in \cite{chung}.

\section{Analysis of the Particle Spectrum}

\begin{figure}[t]
\begin{center}
  \epsfig{file=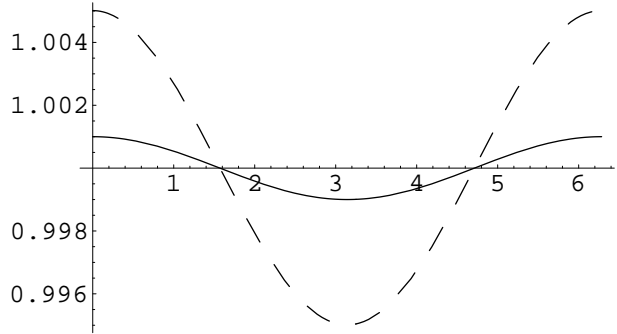,angle=0,width=0.5\linewidth} 
\end{center}
\caption{\it The ratio of the Bogolubov coefficient (\ref{spectrumfinal}) 
over the unperturbed one $u=0$ (\ref{inflationbog})  
versus the angle $\theta \in [0,2\pi]$ for two characteristic 
momentum scales  $k_1$ (dashed)  and $k_2$ (continuous),
with $k_2 >> k_1$ (in units of $H$ during inflation).}
\label{spectrumgraph}
\end{figure}

In this section we will explore some of the possible physical
consequences of the spectrum (\ref{spectrumfinal}), incorporating 
the back-reaction effects of the D-particle recoil onto the inflationary
space-time. First of all, the spectrum is {\it non thermal}, due to 
the presence of the ${\rm cos\theta}$ term. This will 
induce anisotropies into the spectrum of the emitted Dark matter 
particles (c.f. figure \ref{spectrumgraph}), which, in principle, may be 
observable.

As discussed in \cite{chung}, and is valid also in our case, 
the produced particles will be cosmologically abundant, and thus relevant,
i.e. will have densities of order one, if and only if their masses
are {\it at least} 
of order of 
the Hubble parameter during inflation,
$H \sim 10^{-5}M_P$.
Thus, as suggested in \cite{chung}, such heavy particles 
can be candidates for Dark matter. 
This conclusion survives in our case, given that the bulk effect 
of particle production is still due to the exponential expansion of the 
Universe (the distortions due to D-particle recoil are sub-dominant
effects). 
Therefore the anisotropies in our case, inferred from (\ref{spectrumfinal}),  
will in general imply anisotropies
in the superheavy dark matter spectrum, which could have 
phenomenological significance.

The effects leading to (\ref{spectrumfinal}) stem from  
single scattering events between a string and a D-particle.
In general, one encounters a distribution of D-particles in 
the Early Universe, characterized by a density $\rho_{\scriptscriptstyle D}$. 
Assuming for simplicity uniform densities, one can incorporate 
the effects of such ensembles of D-particles by 
replacing $u$ in the metric (\ref{recoilmetricfl}) by 
$\rho_{\scriptscriptstyle D} {\overline u}$, where 
the bar indicates average quantities
over the ensemble. In addition, one should 
integrate (\ref{spectrumfinal})
over all possible angles $\theta$. 
Setting 
\begin{equation}
z_k = a_0^2 |u| H^{-1} \sqrt{k^2 + A M^2} \,,\quad
{\bar z}_k \equiv a_0^2 |{\overline u}|H^{-1}\sqrt{k^2 + A M^2} \,,
\end{equation}
the integral over angles, for uniform 
D-particle densities and distributions, can be performed 
analytically~\cite{integrals}, and yields:
\begin{eqnarray}
|\beta _k^{({\overline u})}|^2 &=& 
|\beta _k|^2 
\int_0^{2\pi}  d\theta \,
{\rm exp}\left( -{\bar z}_k \rho_{\scriptscriptstyle D} \cos\theta\right)  
=\nonumber \\
&& 2\pi |\beta _k|^2 I_0({\bar z}_k\rho_{\scriptscriptstyle D} )~,
\label{spectrumangular}
\end{eqnarray}
where $|\beta _k|^2$ is the unperturbed inflation spectrum 
(\ref{inflationbog}) of \cite{chung}, and 
$I_0(z)$ is the modified Bessel function, which in our case 
can be expressed 
in terms of the Bessel function ${\cal J}_0(z)$ as: $I_0(z)= 
{\cal J}_0(e^{i\frac{\pi}{2}}z)$. 
For small arguments, as is the case above due to the smallness of 
${\overline u}$, we may use the perturbative expansion~\cite{integrals}:
\begin{equation}
I_0(x) = \sum_{k=0}^{\infty} \frac{x^{2k}}{2^{2k}k!\Gamma (k+1)} ,
\end{equation}
which to order ${\overline u}^2$ yields from (\ref{spectrumangular}):
\begin{eqnarray}
|\beta _k^{({\overline u})}|^2 
= |\beta _k|^2 \left( 1 + \frac{1}{4} \rho_{\scriptscriptstyle D}^2 {\bar z}_k^2 + \dots \right)
\label{spectrumangularfinal}
\end{eqnarray}

Another quantity of interest is the total number of emitted particles,
from a single scattering event of a string colliding with a D-particle.
This is 
obtained upon integrating (\ref{spectrumfinal}) over momentum scale vectors
$d^3 k = k^2 dk\, {\rm sin}\theta\, d\theta \,d\phi $, where we selected the 
axis of the recoil velocity ${\vec u}$ to coincide with the $k_z$ axis.
The result is (c.f. (\ref{creation})):
\begin{eqnarray}
n^{u}_{X(t)} = \int_0^{\infty}\frac{d^3k}{(2\pi)^3 a^3(t)}  
|\beta _k^{({u})}|^2 = \int_0^\infty \frac{k^2 dk}{2\pi^2a^3(t)}\frac{1}{z_k}{\rm 
sinh}(z_k)|\beta_k|^2~;
\end{eqnarray}
Expanding for small $u$, and keeping terms up to order $u^2$, we obtain:  
\begin{equation} 
n^{u}_{X(t)} - n_{X(t)} \simeq \frac{1}{6} a_0^4 |u|^2 H^{-2} \int_0^\infty dk k^2 (k^2 + A M^2)  
|\beta_k|^2 \simeq \frac{A}{6} a_0^4 |u|^2 \frac{M^2}{H^2} n_X(t) 
\label{ucreation}
\end{equation}
where $n_X(t)$ is the unperturbed total particle 
density ($u=0$) (\ref{density}), and  
in the last (approximate) equality on the right-hand-side 
of (\ref{ucreation})  we have considered for simplicity 
the case of low momenta $k << M$. 

\section{Conclusions and Outlook} 

In this article we have presented a brane-inspired model 
of anisotropies (\ref{spectrumfinal}) 
in the spectrum of superheavy (masses $M > 10^{13} $ GeV) 
Dark matter particles.
This may also imply an anisotropic spectrum of ultra-high-energy
cosmic rays, should the latter be the product of decay of such
superheavy particles, according to some scenaria~\cite{kryptons}. 
Although at present there seems to be no experimental evidence 
for such anisotropies~\cite{uhecranis}, however, the experimental
situation is far from being conclusive, mainly due to the 
scarcity of the ultra-high-energy cosmic ray events available to date.
The situation is about to be improved significantly in the 
near future, whereby the launch of experiments such as the Piere Auger 
will improve the available statistics by several orders of magnitude
(provided the ultra-high-energy cosmic ray events are actually there).

On the other hand, if anisotropies are not detected in such improved facilities,
one should be able to place strict upper bounds 
on parameters of models like the one presented here, 
for instance on the quantity ${\overline u}\rho_D$ entering the spectrum 
(\ref{spectrumfinal}), (\ref{spectrumangularfinal}).
Such bounds, if combined with progress in the theoretical
understanding of D-brane theory and cosmology, may be 
instrumental in seriously constraining non-critical string theory models 
of the type employed here.

We have arrived at the above conclusions using 
perturbative $\sigma$-model conformal field theory methods,
which we expect to be 
applicable at some regime of the parameters
of an inflationary Universe. 
We are fully aware of the possibility that non-perturbative string 
Physics may be responsible for the inflationary background, on which 
our calculations have been based, and we have therefore also provided 
arguments on how one could tackle the problem using perturbative $\sigma$-model
methods, at the cost, however, of departing from critical string theory.
We demonstrated the mathematical consistency of such an approach,
at least within a given time range.
In fact, the non-critical string theory approach to 
inflation has many phenomenologically desirable features, 
including the possibility of accelerating phases of 
the Universe~\cite{papant}, 
and thus non-zero dark energy components
in agreement with 
standard (TeV scale) supersymmetry breaking scenaria~\cite{grav2}.

There are many avenues for improving the results presented in this 
article. First of all, one should use realistic string Universe brane 
models, by taking into account the standard model fields on the brane
and/or including supersymmetry and its breaking in a consistent way
(e.g. through compactification of extra dimensions in five-branes 
on magnetized tori, as in \cite{grav2}: such models are known to 
have instabilities, but in a cosmological context this may be a 
desirable feature that needs to be explored). 
Second, one should improve on the effective field theory 
computation of particle production presented in this article by
using as interpolating metrics realistic space-time metrics
obtained from the above non-critical string models. At present
the situation is only known numerically~\cite{papant}, 
but we may not be far from 
obtaining some analytic results, 
given the increasing recent interest 
towards issues like inflationary physics in the context 
of string theory from various viewpoints~\cite{martinec,gubser}.

Finally, in this context, we stress the preliminary conclusions of 
\cite{gubser}, on the potentially important 
r\^ole of strings at the end of inflation 
towards reheating and other related issues.
It would be interesting to place the last issue in the framework 
of non-critical strings discussed here.
As argued above, such a framework  
appears to provide a 
viable treatment of inflationary dynamics within 
the context of perturbative string theory, with the potential  
of providing us with 
experimentally testable (in principle) predictions. 

\section*{Acknowledgements} 

This paper is dedicated to the memory of Ian I. Kogan, collaborator
and friend. The work of N.E.M. is partially 
supported by a visiting professorship
at the University of Valencia (Spain),
Department of Theoretical Physics, and by the European Union 
(contract HPRN-CT-2000-00152).  
The work of J.P. is supported by the Grants BFM2002-000568, FPA2002-00612, 
and GV01-94. 
\newpage

\section*{Appendix: Inflation as a Liouville String $\sigma$-model  Background}
 
{\bf A Concrete Example of Non-critical Strings: Colliding Brane Worlds} 
\paragraph{}
In this Appendix we discuss briefly how an inflationary space-time 
may be derived as a consistent background of a non-critical
string theory. Although the approach is general, 
and can be applied to many non-critical string models, 
for definiteness
we shall concentrate here on one particular case~\cite{grav2}, where the 
non-criticality is induced by the collision of two 
brane worlds. 

In what follows we shall first discuss the basic features
of this scenario, and then proceed to demonstrate explicitly 
the emergence of inflationary space-times from such situations. 
Following \cite{grav2}, we consider two five-branes of type II string theory,
in which the extra two dimensions have been compactified on tori. In one
of the branes (assumed to be the hidden (not our) 
world) the torus is magnetized,
with a magnetic field intensity ${\cal H}$. 
Initially our world is compactified on a normal torus, without 
magnetic fields, and the two branes are assumed on collision course
in the bulk, with a small velocity $v \ll 1$.  
The collision produces a non-equilibrium situation, which results in 
electric current transfer from the   hidden to the visible brane. 
This causes the (adiabatic) emergence  of magnetic fields in our world. 

The associated instabilities that accompany such magnetized-tori 
compactifications are not a problem in the context of cosmological 
scenaria we are discussing here. 
In fact, as discussed in \cite{grav2}, the collision may also produce
decompactification of the extra toroidal dimensions, which however 
takes place at a rate much slower than any other rate in the problem.
As discussed in \cite{grav2}, this is important 
for guaranteeing an asymptotic equilibrium situation and a proper 
definition of an S-matrix for the stringy excitations on the observable 
world. 

The collision of the two branes implies, for a short period after 
it, when the branes are at most a few string scales apart,
the exchange of open string excitations stretching between the branes 
with their ends attached on them.
As argued in \cite{grav2}, the exchange of such pairs of 
open strings in the type II string theory considered in that work
result in an excitation energy on the visible world. 
The latter may be estimated
by computing the corresponding scattering amplitude
of the two branes, using string-theory world-sheet methods~\cite{bachas}.
Essentially the time integral for the relevant potential yields
the scattering amplitude. 
Such estimates involve the computation of appropriate world-sheet 
annulus diagrams, due to the existence of open string pairs in 
type II string theory. This implies the presence of ``spin factors''
as proportionality constants 
in the scattering amplitude, which are expressed in terms of 
Jacobi $\Theta$ functions. For small brane velocities $v \ll 1$ 
we are considering in our case, the appropriate spin structures
start at {\it quartic order } in $v$, as a result of 
mathematical properties of the Jacobi functions~\cite{bachas}.
This in turn implies~\cite{grav2} that the resulting excitation 
energy on the brane world is of order $V = {\cal O}(v^4)$, which 
may be thought of as an initial (approximately constant)
value of a {\it supercritical}  central-charge deficit for the non-critical 
$\sigma$-model describing stringy excitations on the 
observable world after the collision:
\begin{equation}
Q^2 = {\cal O}\left( v^4 \right) > 0
\label{initialdeficit}
\end{equation} 
The supercriticality of the model is essential~\cite{aben} 
for a time-like signature of the Liouville mode.

As we discuss below, such constant $Q^2$ can produce inflation 
with a scale factor varying exponentially with the Robertson-Walker 
time: $a(t) \sim e^{Q t}$ (up to proportionality numerical constants in the 
exponent, of order one, which can be fixed by normalization, c.f. below). 
The duration of the inflationary era is therefore of order $t_* \sim 1/Q$,
which matches the conventional cosmology 
estimates of $10^9t_P$ for non-relativistic brane velocities 
$v^2 \sim 10^{-9}$. These small velocities 
are consistent with our perturbative 
$\sigma$-model
formalism of recoil~\cite{kogan} we follow here.  

For times long after the collision, the central charge deficit 
is no longer constant but relaxes with time $t$. 
In the approach of 
\cite{grav2} such a relaxation 
has been computed by means 
of logarithmic conformal field theory world-sheet methods~\cite{lcft,kogan}, 
taking into account recoil (in the bulk) of the observable-world brane 
and the identification of target 
time with the (zero mode of the) Liouville field. 
This late-time varying deficit $Q^2(t)$ 
has been identified~\cite{grav2} with a
`quintessence-like' dark energy density component of our world: 
\begin{equation} 
\Lambda (t) \sim \frac{R^2({\cal H}^2 + v^2)^2}{t^2}
\left(\frac{M_s}{M_P}\right)^4M_P^4~,
\label{cosmoconst}
\end{equation} 
where $R$ is the 
compactification radius. In our model, for the validity 
of the $\sigma$-model perturbative formalism the 
constraints  (\ref{relconstr}) apply, which lead naturally
to $M_s \sim 10^{-4}M_P$. 

We next remark that 
the presence of the magnetic field ${\cal H}$ is responsible for a breaking of 
target-space 
supersymmetry~\cite{bachas2}, due to the fact that bosons and 
fermions on the brane worlds couple differently to ${\cal H}$. 
The resulting mass difference between bosonic and fermionic string 
excitations 
for our problem, where the magnetic field 
is turned on adiabatically, is found to be~\cite{grav2}:
\begin{equation} 
\Delta m^2_{\rm string} \sim 2{\cal H}{\rm cosh}\left(\epsilon \varphi + 
\epsilon t\right)\Sigma_{45}
\label{masssplit}
\end{equation} 
where $\Sigma_{45}$ is a standard spin operator on the plane of the torus,
and $\epsilon \to 0^+$ is the regulating parameter 
of the Heaviside operator (\ref{regulator})
appearing in the D-brane recoil formalism~\cite{kogan}.  
>From (\ref{masssplit}) we observe that the formalism selects dynamically a 
Liouville mode which flows opposite to the target time $\varphi = -t$,
as a result of minimization of the effective field-theoretic potential 
of 
the various stringy excitations. 

By choosing appropriately ${\cal H}$ we may thus arrange for the 
supersymmetry breaking scale to be of order of a few TeV. 
It turns out that the so-chosen magnetic field contribution  
is subdominant, as compared with the velocity contribution $v^2 \sim 10^{-9}$,
in the expression for the current era dark energy (\ref{cosmoconst}).  
The model, therefore, is capable of reproducing naturally 
a current value of the dark energy
(i.e. for $t \sim 10^{60}t_P$)
compatible with observations~\cite{supernovae,wmap},
provided one chooses relatively large 
compactification radii $R \sim 10^{17}\ell_P \sim 10^{-18}$ m, which 
are common in modern string theories~\footnote{For models 
where the compactification involves higher-dimensional manifolds
than tori 
a volume factor $R^{n}$, with $n > 2$ the number of extra dimensions, appears 
in (\ref{cosmoconst}), and thus in such cases the compactification 
radii are significantly smaller.}.   
However, 
we cannot claim at this stage that this 
(toy) model is free from fine tuning, since 
the final asymptotic value of the central charge deficit has 
been arranged to vanish, by an appropriate choice of various 
constants appearing in the problem~\cite{grav2}. This  
is required by the assumption  
that our non-critical string system 
relaxes asymptotically in time to a critical string.  
In the complete model, the identification of the Liouville field with 
target time~\cite{emn,kogan} would define the appropriate 
renormalization-group trajectory, which hopefully would pick up
the appropriate asymptotic critical string state  dynamically. This
still remains to be seen analytically
in realistic models, although it has been demonstrated numerically
for some stringy models in \cite{papant}. 
Nevertheless, the current toy example 
is sufficient in providing a non-trivial, and physically relevant, 
concrete example of an inflationary Universe in the context of Liouville
strings, a demonstration of which we now turn to. 
\paragraph{}

\noindent{\bf Inflation from Liouville Strings} 

\paragraph{}
Consider a non-critical $\sigma$-model in metric ($G_{\mu\nu}$), 
antisymmetric tensor ($B_{\mu\nu}$),  
and dilaton backgrounds ($\Phi$), with the following ${\cal O}(\alpha ')$ 
$\beta$-functions ($\alpha '$ the Regge slope)~\cite{tseytlin}:
\begin{eqnarray} 
&& \beta^G_{\mu\nu} = \alpha ' \left( R_{\mu\nu} + 2 \nabla_{\mu} 
\partial_{\nu} \Phi 
 - \frac{1}{4}H_{\mu\rho\sigma}H_{\nu}^{\rho\sigma}\right)~, \nonumber \\
&& \beta^B_{\mu\nu} = \alpha '\left(-\frac{1}{2}\nabla_{\rho} H^{\rho}_{\mu\nu} + 
H^{\rho}_{\mu\nu}\partial_{\rho} \Phi \right)~, \nonumber \\
&& {\tilde \beta}^\Phi = \beta^\Phi - \frac{1}{4}G^{\rho\sigma}\beta^G_{\rho\sigma} = 
\frac{1}{6}\left( C - 26 \right)
\label{bfunctions}
\end{eqnarray} 
where Greek indices are four-dimensional, 
including target-time components $\mu, \nu, ...= 0,1,2,3$
on the D3 brane worlds of ref. \cite{grav2}, 
and 
$H_{\mu\nu\rho}= \partial_{[\mu}B_{\nu\rho]}$ is the field strength. 

We consider the following representation of the four-dimensional 
field strength in terms of a pseudoscalar (axion-like) field $b$: 
\begin{equation}
H_{\mu\nu\rho} = \epsilon_{\mu\nu\rho\sigma}\partial^\sigma b 
\label{axion}
\end{equation}
where $\epsilon_{\mu\nu\rho\sigma}$ is the four-dimensional antisymmetric symbol. 

Choose the linear in time axion background~\cite{aben}: 
\begin{equation} 
b = b(t) = \beta t~, \quad  \beta={\rm constant}
\label{axion2}
\end{equation}
which yields a constant field strength with spatial indices only:
$H_{ijk} = \epsilon_{ijk}\beta$, $H_{0jk}= 0$.  
This implies that this background is a conformal solution of the full 
four-dimensional antisymmetric tensor ${\cal O}(\alpha')$ $\beta$-function. 

We also consider a dilaton background linear in time t \cite{aben}
\begin{equation}
\Phi (t,X) = {\rm const} + ({\rm const})' t 
\label{constdil}
\end{equation}
This background does not contribute to the antisymmetric tensor and 
metric $\beta$-functions of (64).

Suppose, now, that only the metric is a non conformal background, due to 
some initial quantum fluctuations or catastrophic events, like the collision
of two brane worlds discussed above and in \cite{grav}, 
which result in the presence
of an initial central charge deficit $Q^2$ (\ref{initialdeficit}), 
constant at early stages after the 
collision. Let 
\begin{equation} 
G_{ij} = e^{\kappa \varphi + Hct}\eta_{ij}~, \quad G_{00}=e^{\kappa '\varphi 
+ Hct}\eta_{00}
\label{metricinfl}
\end{equation}
where $t$ is the target time, $\varphi$ is the Liouville mode, 
$\eta_{\mu\nu}$ is the four-dimensional Minkowski metric, 
and $\kappa, \kappa ', c$ are constants to be determined. 

According to the discussion in the text, following (\ref{liouveq}),  
we require:
\begin{equation}
Q = -3H 
\label{qh}
\end{equation}
which partially stems from~\cite{kogan1,emn} 
\begin{equation}
\varphi = -t
\label{liouvtime}
\end{equation}
This restriction will be imposed dynamically~\cite{grav,emn}
only at the end of our computations. Initially, one should treat
$\varphi, t$ as independent target space components. 

The Liouville-dressing induces~\cite{ddk} $\sigma$-model terms of the form 
$\int_{\Sigma} R^{(2)} Q \varphi$, where $R^{(2)}$ is the world-sheet curvature.
Such terms lead to non-trivial contributions to the dilaton background in the 
(D+1)-dimensional space-time $(\varphi,t,X^i)$:
\begin{equation}
\Phi (\varphi,t,X^i) = Q \,\varphi + ({\rm const})' t + {\rm const}
\end{equation}
Upon choosing $({\rm const})'=Q$, we observe that (70) implies a 
constant dilaton background.

We now consider the Liouville-dressed equations (\ref{liouveq}) 
for the $\beta$-functions of the metric and antisymmetric 
tensor fields (\ref{bfunctions}). For constant dilatons 
that we assume, the dilaton equation yields no independent information, 
apart from expressing the dilaton $\beta$ function in terms of the 
central charge deficit as usual. For the axion background (\ref{axion2}) 
only the metric yields a non-trivial constraint (we work in units of 
$\alpha ' =1$ for convenience):
\begin{equation} 
G''_{ij} + QG'_{ij} = -R_{ij} + \frac{1}{2}\beta^2 G_{ij}
\end{equation}
where the prime indicates differentiation with respect to the 
(world-sheet zero mode of the) Liouville mode $\varphi$, and $R_{ij}$ is the 
(non-vanishing) Ricci tensor of the (non-critical) $\sigma$-model 
with coordinates $(t,{\vec x})$: 
$R_{00}=0~, R_{ij}=\frac{c^2H^2}{2}e^{(\kappa - \kappa ')\varphi}$.

One should also take into account the temporal 
($t$) equation for the metric tensor (for the antisymmetric backgrounds 
this is identically zero): 
\begin{equation}
G''_{00} + QG'_{00} = -R_{00} = 0
\label{tempgrav}
\end{equation}
where the vanishing of the Ricci tensor stems from the 
specific form of the background (\ref{metricinfl}).

Moreover we seek metric backgrounds in Robertson-Walker inflationary 
form (de Sitter): 
\begin{equation}
G_{00}=-1~, \quad G_{ij}=e^{2Ht}\eta_{ij}
\label{desittermetric}
\end{equation}
Then, 
from (\ref{desittermetric}), (\ref{metricinfl}),
(\ref{constdil}),(\ref{axion2}) and (\ref{qh}), 
and imposing (\ref{liouvtime}) at the end,   
we do observe that there is a 
consistent solution with: 
\begin{equation}
Q = -3H = - \kappa ',~c=3,~\kappa = H,~\beta^2 = 5H^2
\label{solution}
\end{equation}

It is in such backgrounds that we consider the (back reaction) 
effects of recoiling D-particles, by employing again 
Liouville dressing techniques and the identification~\cite{grav,kogan} 
(\ref{liouvtime}). As discussed in the text, we consider 
the effects only for times $0 < Ht << 1$ for the validity of our 
$\sigma$-model perturbation theory. As a consistency check
of our approach, we find that the identification (\ref{liouvtime}),
with opposite flows of Liouville and target time
fields, implies the decay of the back reaction effects as the time elapses
(c.f. (\ref{distorted}),(\ref{newcoordinates})).


\begin{thebibliography}{99}

\bibitem{supernovae} B.~P.~Schmidt {\it et al.},
%``The High-Z Supernova Search: Measuring Cosmic Deceleration and Global Cur vature of the Universe Using Type Ia Supernovae,''
Astrophys.\ J.\  {\bf 507} (1998) 46
[arXiv:astro-ph/9805200];
%%CITATION = ASTRO-PH 9805200;%%
S.~Perlmutter {\it et al.}  [Supernova Cosmology Project Collaboration],
%``Measurements of Omega and Lambda from 42 High-Redshift Supernovae,''
Astrophys.\ J.\  {\bf 517} (1999) 565
[arXiv:astro-ph/9812133];
%%CITATION = ASTRO-PH 9812133;%%
J.~P.~Blakeslee {\it et al.},
%``Discovery of Two Distant Type Ia Supernovae in the Hubble Deep Field North with the Advanced Camera for Surveys,''
Astrophys.\ J.\  {\bf 589} (2003) 693
[arXiv:astro-ph/0302402];
%%CITATION = ASTRO-PH 0302402;%%
A.~G.~Riess {\it et al.},
%``The Farthest Known Supernova: Support for an Accelerating Universe and a Glimpse of the Epoch of Deceleration,''
Astrophys.\ J.\  {\bf 560} (2001) 49
[arXiv:astro-ph/0104455].
%%CITATION = ASTRO-PH 0104455;%%



\bibitem{wmap} C.~L.~Bennett {\it et al.},
%``First Year Wilkinson Microwave Anisotropy Probe (WMAP) Observations: Preliminary Maps and Basic Results,''
arXiv:astro-ph/0302207;
%%CITATION = ASTRO-PH 0302207;%%
D.~N.~Spergel {\it et al.},
%``First Year Wilkinson Microwave Anisotropy Probe (WMAP) Observations: Determination of Cosmological Parameters,''
arXiv:astro-ph/0302209;
%%CITATION = ASTRO-PH 0302209;%%
H.~V.~Peiris {\it et al.},
%``First year Wilkinson Microwave Anisotropy Probe (WMAP) observations:  Implications for inflation,''
arXiv:astro-ph/0302225.
%%CITATION = ASTRO-PH 0302225;%%




\bibitem{bellido} For recent reviews, in light of WMAP data, see:
J.~Garcia-Bellido,
%``The evolution of the universe,''
arXiv:hep-ph/0303153;
%%CITATION = HEP-PH 0303153;%%
J.~R.~Ellis,
%``Dark matter and dark energy: Summary and future directions,''
arXiv:astro-ph/0304183 
%%CITATION = ASTRO-PH 0304183;%%
and 
arXiv:astro-ph/0305038, and references therein.  
%``Particle physics and cosmology,''
%%CITATION = ASTRO-PH 0305038;%%





\bibitem{inflation} A.~H.~Guth,
%``The Inflationary Universe: A Possible Solution To The Horizon And Flatness Problems,''
Phys.\ Rev.\ D {\bf 23} (1981) 347;
%%CITATION = PHRVA,D23,347;%%
A.~D.~Linde,
%``A New Inflationary Universe Scenario: A Possible Solution Of The Horizon, Flatness, Homogeneity, Isotropy And Primordial Monopole Problems,''
Phys.\ Lett.\ B {\bf 108} (1982) 389.
%%CITATION = PHLTA,B108,389;%%



\bibitem{wmapinflaton} W.~H.~Kinney, E.~W.~Kolb, A.~Melchiorri and A.~Riotto,
%``WMAPping inflationary physics,''
arXiv:hep-ph/0305130;
%%CITATION = HEP-PH 0305130;%%


\bibitem{carroll}S.~M.~Carroll,
%``The cosmological constant,''
Living Rev.\ Rel.\  {\bf 4} (2001) 1
[arXiv:astro-ph/0004075], and references therein.
%%CITATION = ASTRO-PH 0004075;%%


\bibitem{strings} J.~Polchinski,
{\it String Theory.} Vols. 1,2 (Cambridge University Press (U.K.) 1998).
%\href{http://www.slac.stanford.edu/spires/find/hep/www?irn=4634802}{SPIRES entry}




\bibitem{smatrix} S.~Hellerman, N.~Kaloper and L.~Susskind,
%``String theory and quintessence,''
JHEP {\bf 0106} (2001) 003
[arXiv:hep-th/0104180];
%%CITATION = HEP-TH 0104180;%%
W.~Fischler, A.~Kashani-Poor, R.~McNees and S.~Paban,
%``The acceleration of the universe, a challenge for string theory,''
JHEP {\bf 0107} (2001) 003
[arXiv:hep-th/0104181];
%%CITATION = HEP-TH 0104181;%%
J.~R.~Ellis, N.~E.~Mavromatos and D.~V.~Nanopoulos,
%``String theory and an accelerating universe,''
arXiv:hep-th/0105206.
%%CITATION = HEP-TH 0105206;%%


\bibitem{papant} G.~A.~Diamandis, B.~C.~Georgalas, N.~E.~Mavromatos, E.~Papantonopoulos and I.~Pappa,
%``Cosmological evolution in a type-0 string theory,''
Int.\ J.\ Mod.\ Phys.\ A {\bf 17} (2002) 2241
[arXiv:hep-th/0107124]; 
%%CITATION = HEP-TH 0107124;%%
G.~A.~Diamandis, B.~C.~Georgalas, N.~E.~Mavromatos and E.~Papantonopoulos,
%``Acceleration of the universe in type-0 non-critical strings,''
Int.\ J.\ Mod.\ Phys.\ A {\bf 17} (2002) 4567
[arXiv:hep-th/0203241].
%%CITATION = HEP-TH 0203241;%%




\bibitem{branecosm} P.~Binetruy, C.~Deffayet and D.~Langlois,
%``Non-conventional cosmology from a brane-universe,''
Nucl.\ Phys.\ B {\bf 565} (2000) 269
[arXiv:hep-th/9905012];
%%CITATION = HEP-TH 9905012;%%
For a comprehensive review see: D.~Langlois,
%``Brane cosmology: An introduction,''
Prog.\ Theor.\ Phys.\ Suppl.\  {\bf 148} (2003) 181
[arXiv:hep-th/0209261], and references therein.
%%CITATION = HEP-TH 0209261;%%



\bibitem{grav2} E.~Gravanis and N.~E.~Mavromatos,
%``Vacuum energy and cosmological supersymmetry breaking in brane worlds,''
Phys.\ Lett.\ B {\bf 547} (2002) 117
[arXiv:hep-th/0205298]; see also
%%CITATION = HEP-TH 0205298;%%
N.~E.~Mavromatos,
{\it Quantum gravity, cosmology, (Liouville) strings and Lorentz invariance}, 
sec 4.2 [arXiv:hep-th/0210079] (Proc. {\it Beyond the Desert 2002} (Oulu Finland, June 2002), in press).  
%%CITATION = HEP-TH 0210079;%%

\bibitem{ekpyrotic} J.~Khoury, B.~A.~Ovrut, P.~J.~Steinhardt and N.~Turok,
%``The ekpyrotic universe: Colliding branes and the origin of the hot big  bang,''
Phys.\ Rev.\ D {\bf 64} (2001) 123522
[arXiv:hep-th/0103239].
%%CITATION = HEP-TH 0103239;%%



\bibitem{linde} R.~Kallosh, L.~Kofman, A.~D.~Linde and A.~A.~Tseytlin,
%``BPS branes in cosmology,''
Phys.\ Rev.\ D {\bf 64} (2001) 123524
[arXiv:hep-th/0106241];
%%CITATION = HEP-TH 0106241;%%
A.~Linde,
%``Inflationary theory versus ekpyrotic / cyclic scenario,''
arXiv:hep-th/0205259.
%%CITATION = HEP-TH 0205259;%%



\bibitem{seiberg} J.~Khoury, 
B.~A.~Ovrut, N.~Seiberg, P.~J.~Steinhardt and N.~Turok,
%``From big crunch to big bang,''
Phys.\ Rev.\ D {\bf 65} (2002) 086007
[arXiv:hep-th/0108187].
%%CITATION = HEP-TH 0108187;%%




\bibitem{ddk} F.~David,
\newblock Mod. Phys. Lett. {\bf A3}, 1651 (1988);
%%CITATION = MPLAE,A3,1651;%%
J.~Distler and H.~Kawai,
\newblock Nucl. Phys. {\bf B321}, 509 (1989);
%%CITATION = NUPHA,B321,509;%%
see also: N.~E.~Mavromatos and J.~L.~Miramontes,
%``Regularizing The Functional Integral In 2-D Quantum Gravity,''
Mod.\ Phys.\ Lett.\ A {\bf 4} (1989) 1847;
%%CITATION = MPLAE,A4,1847;%%
E.~D'Hoker and P.~S.~Kurzepa,
%``2-D Quantum Gravity And Liouville Theory,''
Mod.\ Phys.\ Lett.\ A {\bf 5} (1990) 1411;
%%CITATION = MPLAE,A5,1411;%%
I.~R.~Klebanov, I.~I.~Kogan and A.~M.~Polyakov,
%``Gravitational dressing of renormalization group,''
Phys.\ Rev.\ Lett.\  {\bf 71} (1993) 3243
[arXiv:hep-th/9309106].
%%CITATION = HEP-TH 9309106;%%

\bibitem{emn} J.~R.~Ellis, N.~E.~Mavromatos and D.~V.~Nanopoulos,
%``String theory modifies quantum mechanics,''
Phys.\ Lett.\ B {\bf 293} (1992) 37
[arXiv:hep-th/9207103], for a review see: 
%``A Liouville string approach to microscopic time and cosmology,''
arXiv:hep-th/9311148, {\it Erice lectures}, 
Sub-nuclear series, Vol. 31, 1 (World Sci. 1994).
%%CITATION = HEP-TH 9311148;%%

\bibitem{kogan1}I.~I.~Kogan,
%``Time As Liouville Field And World Sheet Rg Equations As Evolution Equations,''
%\href{http://www.slac.stanford.edu/spires/find/hep/www?irn=4177460}{SPIRES entry}
{\it Prepared for Particles \& Fields 91: Meeting of the Division of Particles \& Fields of the APS, Vancouver, Canada, 18-22 Aug 1991};
%``Regularization Of The Area Integration And Imaginary Parts Of The Correlation Functions In Liouville Theory,''
Phys.\ Lett.\ B {\bf 265} (1991) 269.
%%CITATION = PHLTA,B265,269;%%


\bibitem{lcft} Partial list of references: V.~Gurarie,
%``Logarithmic operators in conformal field theory,''
Nucl.\ Phys.\ B {\bf 410} (1993) 535
[arXiv:hep-th/9303160];
%%CITATION = HEP-TH 9303160;%%
M.~A.~Flohr,
%``On Modular Invariant Partition Functions of Conformal Field Theories with Logarithmic Operators,''
Int.\ J.\ Mod.\ Phys.\ A {\bf 11} (1996) 4147
[arXiv:hep-th/9509166] and 
%%CITATION = HEP-TH 9509166;%%
%``Bits and pieces in logarithmic conformal field theory,''
arXiv:hep-th/0111228;
%%CITATION = HEP-TH 0111228;%%
A.~Bilal and I.~I.~Kogan,
%``On gravitational dressing of 2-D field theories in chiral gauge,''
Nucl.\ Phys.\ B {\bf 449} (1995) 569
[arXiv:hep-th/9503209].
%%CITATION = HEP-TH 9503209;%%
J.~S.~Caux, I.~I.~Kogan and A.~M.~Tsvelik,
%``Logarithmic Operators and Hidden Continuous Symmetry in Critical Disordered Models,''
Nucl.\ Phys.\ B {\bf 466} (1996) 444
[arXiv:hep-th/9511134];
%%CITATION = HEP-TH 9511134;%%
M.~R.~Gaberdiel and H.~G.~Kausch,
%``Indecomposable Fusion Products,''
Nucl.\ Phys.\ B {\bf 477} (1996) 293
[arXiv:hep-th/9604026];
%%CITATION = HEP-TH 9604026;%%
M.~R.~Rahimi Tabar, A.~Aghamohammadi and M.~Khorrami,
%``The logarithmic conformal field theories,''
Nucl.\ Phys.\ B {\bf 497} (1997) 555
[arXiv:hep-th/9610168];
%%CITATION = HEP-TH 9610168;%%
I.~I.~Kogan and A.~Lewis,
%``Origin of logarithmic operators in conformal field theories,''
Nucl.\ Phys.\ B {\bf 509} (1998) 687
[arXiv:hep-th/9705240].
%%CITATION = HEP-TH 9705240;%%
I.~I.~Kogan,
%``Singletons and logarithmic CFT in AdS/CFT correspondence,''
Phys.\ Lett.\ B {\bf 458} (1999) 66
[arXiv:hep-th/9903162].
%%CITATION = HEP-TH 9903162;%%
I.~I.~Kogan and J.~F.~Wheater,
%``Boundary logarithmic conformal field theory,''
Phys.\ Lett.\ B {\bf 486} (2000) 353
[arXiv:hep-th/0003184].
%%CITATION = HEP-TH 0003184;%%
I.~I.~Kogan and A.~Nichols,
%``SU(2)0 and OSp(2$|$2)(-2) WZNW models: Two current algebras, one  logarithmic CFT,''
Int.\ J.\ Mod.\ Phys.\ A {\bf 17} (2002) 2615
[arXiv:hep-th/0107160;
%%CITATION = HEP-TH 0107160;%%
Y.~Ishimoto,
%``Boundary states in boundary logarithmic CFT,''
Nucl.\ Phys.\ B {\bf 619} (2001) 415
[arXiv:hep-th/0103064].
%%CITATION = HEP-TH 0103064;%%
J.~Fjelstad, J.~Fuchs, S.~Hwang, A.~M.~Semikhatov and I.~Y.~Tipunin,
%``Logarithmic conformal field theories via logarithmic deformations,''
Nucl.\ Phys.\ B {\bf 633} (2002) 379
[arXiv:hep-th/0201091];
%%CITATION = HEP-TH 0201091;%%
N.~E.~Mavromatos and R.~J.~Szabo,
%``The Neveu-Schwarz and Ramond algebras of logarithmic superconformal  field theory,''
JHEP {\bf 0301} (2003) 041
[arXiv:hep-th/0207273];
%%CITATION = HEP-TH 0207273;%%
I.~Bakas and K.~Sfetsos,
%``PP-waves and logarithmic conformal field theories,''
Nucl.\ Phys.\ B {\bf 639} (2002) 223
[arXiv:hep-th/0205006];
%%CITATION = HEP-TH 0205006;%%
K.~Sfetsos,
%``The exact description of NS5-branes in the Penrose limit,''
arXiv:hep-th/0305109.
%%CITATION = HEP-TH 0305109;%%




\bibitem{emninfl} J.~R.~Ellis, N.~E.~Mavromatos and D.~V.~Nanopoulos,
%``A String scenario for inflationary cosmology,''
Mod.\ Phys.\ Lett.\ A {\bf 10} (1995) 1685
[arXiv:hep-th/9503162].
%%CITATION = HEP-TH 9503162;%%





\bibitem{chung} D.~J.~Chung,
%``Classical inflation field induced creation of superheavy dark matter,''
Phys.\ Rev.\ D {\bf 67} (2003) 083514
[arXiv:hep-ph/9809489].
%%CITATION = HEP-PH 9809489;%%





\bibitem{kryptons} J.~R.~Ellis, G.~B.~Gelmini, J.~L.~Lopez, D.~V.~Nanopoulos and S.~Sarkar,
%``Astrophysical Constraints On Massive Unstable Neutral Relic Particles,''
Nucl.\ Phys.\ B {\bf 373} (1992) 399;
%%CITATION = NUPHA,B373,399;%%
P.~Gondolo, G.~Gelmini and S.~Sarkar,
%``Cosmic neutrinos from unstable relic particles,''
Nucl.\ Phys.\ B {\bf 392} (1993) 111
[arXiv:hep-ph/9209236].
%%CITATION = HEP-PH 9209236;%%
V.~Berezinsky, M.~Kachelriess and A.~Vilenkin,
%``Ultra-high energy cosmic rays without GZK cutoff,''
Phys.\ Rev.\ Lett.\  {\bf 79} (1997) 4302
[arXiv:astro-ph/9708217].
%%CITATION = ASTRO-PH 9708217;%%
V.~A.~Kuzmin and V.~A.~Rubakov,
%``Ultrahigh-energy cosmic rays: A window on postinflationary reheating  epoch of the universe?,''
Phys.\ Atom.\ Nucl.\  {\bf 61} (1998) 1028
[Yad.\ Fiz.\  {\bf 61} (1998) 1122]
[arXiv:astro-ph/9709187];
%%CITATION = ASTRO-PH 9709187;%%
M.~Birkel and S.~Sarkar,
%``Extremely high energy cosmic rays from relic particle decays,''
Astropart.\ Phys.\  {\bf 9} (1998) 297
[arXiv:hep-ph/9804285];
%%CITATION = HEP-PH 9804285;%%
G.~Gelmini and A.~Kusenko,
%``Unstable superheavy relic particles as a source of neutrinos  responsible for the ultrahigh-energy cosmic rays,''
Phys.\ Rev.\ Lett.\  {\bf 84} (2000) 1378
[arXiv:hep-ph/9908276];
%%CITATION = HEP-PH 9908276;%%
J.~L.~Crooks, J.~O.~Dunn and P.~H.~Frampton,
%``Relic Neutrinos and Z-Resonance Mechanism for Highest-Energy Cosmic Rays,''
Astrophys.\ J.\  {\bf 546} (2001) L1
[arXiv:astro-ph/0002089];
%%CITATION = ASTRO-PH 0002089;%%
S.~Sarkar,
%``Cosmic ray signatures of massive relic particles,''
arXiv:hep-ph/0005256.
%%CITATION = HEP-PH 0005256;%%







\bibitem{uhecranis} L.~A.~Anchordoqui, C.~Hojvat, T.~P.~McCauley, T.~C.~Paul, S.~Reucroft, J.~D.~Swain and A.~Widom,
%``Full-Sky Search for Ultra High Energy Cosmic Ray Anisotropies,''
arXiv:astro-ph/0305158;
%%CITATION = ASTRO-PH 0305158;%%
C.~Hojvat, T.~P.~McCauley, S.~Reucroft and J.~D.~Swain,
%``Numerical Likelihood Analysis of Cosmic Ray Anisotropies,''
arXiv:astro-ph/0305206.
%%CITATION = ASTRO-PH 0305206;%%





\bibitem{intersect} See, for instance: J.~Garcia-Bellido,
%``Inflation from branes at angles,''
arXiv:astro-ph/0306195, and references therein.
%%CITATION = ASTRO-PH 0306195;%%


\bibitem{kogan} I.~I.~Kogan and N.~E.~Mavromatos,
%``World-Sheet Logarithmic Operators and Target Space Symmetries in String Theory,''
Phys.\ Lett.\ B {\bf 375} (1996) 111
[arXiv:hep-th/9512210];
%%CITATION = HEP-TH 9512210;%%
I.~I.~Kogan, N.~E.~Mavromatos and J.~F.~Wheater,
%``D-brane recoil and logarithmic operators,''
Phys.\ Lett.\ B {\bf 387}, 483 (1996)
[arXiv:hep-th/9606102];
%%CITATION = HEP-TH 9606102;%%
J.~R.~Ellis, N.~E.~Mavromatos and D.~V.~Nanopoulos,
%``D Branes from Liouville Strings,''
Int.\ J.\ Mod.\ Phys.\ A {\bf 12}, 2639 (1997)
[arXiv:hep-th/9605046];
%%CITATION = HEP-TH 9605046;%%
N.~E.~Mavromatos and R.~J.~Szabo,
%``Matrix D-brane dynamics, logarithmic operators and quantization of  noncommutative spacetime,''
Phys.\ Rev.\ D {\bf 59}, 104018 (1999)
[arXiv:hep-th/9808124];
%%CITATION = HEP-TH 9808124;%%
%``D-brane dynamics and logarithmic superconformal algebras,''
JHEP {\bf 0110} (2001) 027
[arXiv:hep-th/0106259].
%%CITATION = HEP-TH 0106259;%%

\bibitem{grav} E.~Gravanis and N.~E.~Mavromatos,
%``Higher-Order Logarithmic Conformal Algebras From Robertson-Walker Sigma-Model Backgrounds,''
JHEP {\bf 0206}, 019 (2002).
%%CITATION = JHEPA,0206,019;%%



\bibitem{otto} H.~Dorn and H.~J.~Otto,
%``On T-duality for open strings in general abelian and nonabelian gauge field backgrounds,''
Phys.\ Lett.\ B {\bf 381}, 81 (1996); 
Nucl.\ Phys.\ Proc.\ Suppl.\  {\bf 56B}, 30 (1997); 
J.~Borlaf and Y.~Lozano,
%``Aspects of T-duality in open strings,''
Nucl.\ Phys.\ B {\bf 480}, 239 (1996); 
Y.~Lozano,
%``Duality and canonical transformations,''
Mod.\ Phys.\ Lett.\ A {\bf 11}, 2893 (1996); 
G.~Amelino-Camelia and N.~E.~Mavromatos,
%``T duality for boundary-non-critical strings,''
Phys.\ Lett.\ B {\bf 422}, 101 (1998).

\bibitem{birrell} See, for instance: N.D. 
Birrell and P.C.W. Davies, {\it Quantum 
Field in Curved Space Times} (Cambridge Univ. Press, Cambridge (U.K.) 1982). 



\bibitem{biswas} S.~Biswas, A.~Shaw and P.~Misra,
%``Particle Production In Expanding Space-Time,''
Gen.\ Rel.\ Grav.\  {\bf 34} (2002) 665;
%%CITATION = GRGVA,34,665;%%
S.~Biswas and I.~Chowdhury,
%``The CWKB particle production and classical condensate in de Sitter  spacetime,''
arXiv:gr-qc/0207058.
%%CITATION = GR-QC 0207058;%%



\bibitem{schwarz} S.~Hossenfelder, D.~J.~Schwarz and W.~Greiner,
%``Particle production in time-dependent gravitational fields: The  expanding mass shell,''
Class.\ Quant.\ Grav.\  {\bf 20} (2003) 2337
[arXiv:gr-qc/0210110].
%%CITATION = GR-QC 0210110;%%



\bibitem{aben} I.~Antoniadis, C.~Bachas, J.~R.~Ellis and D.~V.~Nanopoulos,
%``Cosmological String Theories And Discrete Inflation,''
Phys.\ Lett.\ B {\bf 211} (1988) 393;
%%CITATION = PHLTA,B211,393;%%
%``An Expanding Universe In String Theory,''
Nucl.\ Phys.\ B {\bf 328} (1989) 117;
%%CITATION = NUPHA,B328,117;%%
%``Comments On Cosmological String Solutions,''
Phys.\ Lett.\ B {\bf 257} (1991) 278.
%%CITATION = PHLTA,B257,278;%%







\bibitem{fischler} W.~Fischler and L.~Susskind,
%``Dilaton Tadpoles, String Condensates And Scale Invariance,''
Phys.\ Lett.\ B {\bf 171} (1986) 383;
%%CITATION = PHLTA,B171,383;%%
%``Dilaton Tadpoles, String Condensates And Scale Invariance. 2,''
Phys.\ Lett.\ B {\bf 173} (1986) 262.
%%CITATION = PHLTA,B173,262;%%


\bibitem{coleman} S. Coleman, {\it Aspects of Symmetry}, pp. 278 sec. 2.4 (Cambridge 
University Press 1985). 


\bibitem{zam} A.~B.~Zamolodchikov,
%``'Irreversibility' Of The Flux Of The Renormalization Group In A 2-D Field Theory,''
JETP Lett.\  {\bf 43} (1986) 730
[Pisma Zh.\ Eksp.\ Teor.\ Fiz.\  {\bf 43} (1986) 565].
%%CITATION = JTPLA,43,730;%%






\bibitem{winst} J.~R.~Ellis, P.~Kanti, N.~E.~Mavromatos, D.~V.~Nanopoulos and E.~Winstanley,
%``Decoherent scattering of light particles in a D-brane background,''
Mod.\ Phys.\ Lett.\ A {\bf 13} (1998) 303
[arXiv:hep-th/9711163].
%%CITATION = HEP-TH 9711163;%%



\bibitem{gubser} S.~S.~Gubser,
%``String production at the level of effective field theory,''
arXiv:hep-th/0305099.
%%CITATION = HEP-TH 0305099;%%


\bibitem{integrals} I.S. Gradshteyn and I.M. Ryzhik, {\it Table of Integrals,
Series and Products}, pp. 517, sec 3.915(4) (ed. by A. Jeffrey, Academic Press Inc., New York 1994).


\bibitem{martinec} E.~J.~Martinec,
%``The annular report on non-critical string theory,''
arXiv:hep-th/0305148;
%%CITATION = HEP-TH 0305148;%%
B.~C.~Da Cunha and E.~J.~Martinec,
%``Closed string tachyon condensation and worldsheet inflation,''
arXiv:hep-th/0303087;
%%CITATION = HEP-TH 0303087;%%
A.~Strominger and T.~Takayanagi,
%``Correlators in timelike bulk Liouville theory,''
arXiv:hep-th/0303221;
%%CITATION = HEP-TH 0303221;%%
N.~R.~Constable and F.~Larsen,
%``The rolling tachyon as a matrix model,''
JHEP {\bf 0306} (2003) 017
[arXiv:hep-th/0305177].
%%CITATION = HEP-TH 0305177;%%


\bibitem{bachas} C.~Bachas,
%``D-brane dynamics,''
Phys.\ Lett.\ B {\bf 374} (1996) 37
[arXiv:hep-th/9511043].
%%CITATION = HEP-TH 9511043;%%

\bibitem{bachas2} C.~Bachas,
%``A Way to break supersymmetry,''
arXiv:hep-th/9503030.
%%CITATION = HEP-TH 9503030;%%

\bibitem{tseytlin} R.~R.~Metsaev and A.~A.~Tseytlin,
%``Order Alpha-Prime (Two Loop) Equivalence Of The String Equations Of Motion And The Sigma Model Weyl Invariance Conditions: Dependence On The Dilaton And The Antisymmetric Tensor,''
Nucl.\ Phys.\ B {\bf 293} (1987) 385.
%%CITATION = NUPHA,B293,385;%%





\end{thebibliography}
\end{document}